\documentclass{aa}  
\usepackage{graphicx}
\usepackage{longtable}
\usepackage{lscape}
\usepackage{color}
\usepackage{ulem}    
\usepackage{times}
\usepackage{txfonts}

\def\msun{M$_{\odot}$}

\newcommand{\cmcub}{~cm$^{-3}$}
\newcommand{\kms}{~km~s$^{-1}$}

\newcommand{\Hb}{\ifmmode {\rm H}\beta \else H$\beta$\fi}

\newcommand{\hii}{H\,{\sc ii}}
\newcommand{\hei}{He\,{\sc i}}
\newcommand{\Hei}{He\,{\sc i}\,$\lambda$5876}

\newcommand{\Heii}{He\,{\sc ii}\,$\lambda$4686}

\newcommand{\Nii}{[N\,{\sc ii}]\,$\lambda$6584}

\newcommand{\nii}{[N\,{\sc ii}]}

\newcommand{\oi}{[O\,{\sc i}]}
\newcommand{\Oii}{[O\,{\sc ii}]\,$\lambda$3727}

\newcommand{\Oiii}{[O\,{\sc iii}]\,$\lambda$5007}

\newcommand{\oiii}{[O\,{\sc iii}]}

\newcommand{\Neiii}{[Ne\,{\sc iii}]\,$\lambda$3869}

\newcommand{\Sii}{[S\,{\sc ii}]\,$\lambda$6716,\,$\lambda$6731}
\newcommand{\sii}{[S\,{\sc ii}]}
\newcommand{\Siii}{[S\,{\sc iii}]\,$\lambda$9069}

\newcommand{\Ariii}{[Ar\,{\sc iii}]\,$\lambda$7135}
\newcommand{\ariii}{[Ar\,{\sc iii}]}

\newcommand{\ariv}{[Ar\,{\sc iv}]}

\newcommand{\Oiitoneb}{[O\,{\sc{ii}}]\,$\lambda$7320,7330}

\newcommand{\rOii}{[O\,{\sc ii}]\,$\lambda$3726/3729}

\newcommand{\rOiii}{[O\,{\sc iii}]\,$\lambda$4363/5007}
\newcommand{\rNii}{[N\,{\sc ii}]\,$\lambda$5755/6584}

\newcommand{\rSii}{[S\,{\sc ii}]\,$\lambda$6731/6717}

\newcommand{\rAriv}{[Ar\,{\sc iv}]\,$\lambda$4740/4711}

\newcommand{\Hp}{H$^{+}$}

\newcommand{\Hep}{He$^{+}$}
\newcommand{\Hepp}{He$^{++}$}

\newcommand{\Np}{N$^{+}$}

\newcommand{\Op}{O$^{+}$}
\newcommand{\Opp}{O$^{++}$}

\newcommand{\Nepp}{Ne$^{++}$}
\newcommand{\Sp}{S$^{+}$}

\newcommand{\Arpp}{Ar$^{++}$}

\begin{document}
\title{Planetary nebulae and \hii\ regions in the spiral galaxy NGC\,300}
\subtitle{Clues on the evolution of abundance gradients and on AGB nucleosynthesis
    \thanks{Based on observations collected at the ESO-VLT Observatory,
Paranal, Chile, program 077.B-0430.}}
\author{G. Stasi\'nska  \inst{1}
\and M. Pe\~na  \inst{2}
\and F. Bresolin \inst{3}
\and Y.G.  Tsamis \inst{4}
}

\institute{
LUTH, Observatoire de Paris, CNRS, Universit\'e Paris Diderot; Place Jules Janssen 92190 Meudon, France\\
\email{grazyna.stasinska@obspm.fr}
\and
Instituto de Astronom{\'\i}a, Universidad Nacional Aut\'onoma de M\'exico, Apdo. Postal 70264, M\'ex. D.F., 04510 M\'exico\\
\email{miriam@astro.unam.mx}
\and
Institute for Astronomy, 2680 Woodlawn Drive, Honolulu, HI 96822, USA\\
\email{bresolin@ifa.hawaii.edu}
\and ESO, Karl-Schwarzchild Str. 2, D-85748 Garching bei M{\"u}nchen, Germany\\
\email{ytsamis@eso.org}
}

\offprints{grazyna.stasinska@obspm.fr}

\date{Received    ; accepted      }

\titlerunning{Planetary nebulae and \hii\ regions in the spiral galaxy NGC\,300}
\authorrunning{Stasi\'nska et al.}


\abstract{
We have obtained deep spectra of 26 planetary nebulae (PNe) and 9 compact \hii\ regions in the nearby spiral galaxy NGC 300, and analyzed them together with those of the giant \hii\ regions previously observed. 
We have determined the physical properties of all these objects and their He, N, O, Ne, S and Ar abundances in a consistent way. 

We find that, globally,  compact \hii\ regions have abundance ratios similar to those of giant \hii\ regions, while PNe have systematically larger N/O ratios and similar Ne/O and Ar/O ratios. We demonstrate that the nitrogen enhancement in PNe cannot be only due to second dredge-up in the progenitor stars, since their initial masses are around 2--2.5\,\msun. An extra mixing process is required, perhaps driven by stellar rotation. 

Concerning the  radial abundance distribution,   PNe behave differently from  \hii\ regions: in the central part of the galaxy their average O/H abundance ratio is 0.15 dex smaller.   Their  abundance dispersion at any galactocentric radius is significantly larger than that shown by \hii\ regions and  many of them have  O/H values higher than  \hii\ regions at the same galactocentric distance.  This suggests that not only nitrogen, but also  oxygen is  affected by nucleosynthesis in the PN progenitors, by an amount which depends at least on the stellar rotation velocity and possibly other parameters.   The formal O/H, Ne/H and Ar/He
abundance gradients from PNe are significantly shallower that from \hii\ regions. We  argue that this indicates a steepening of the metallicity gradient in NGC~300 during the last Gyr, rather than an effect of radial stellar motions, although the large observed dispersion  makes this conclusion only tentative.
}
\maketitle
%
\section{Introduction}
\label{sec:intro}

The simultaneous study of the chemical composition of planetary nebulae (PNe) and \hii\ regions in galaxies brings several interesting benefits. While \hii\ regions have  since long been used to probe the present-day chemical composition of the interstellar medium (Peimbert 1975),  the use of PNe, in principle, offers the possibility to study its fossil record, since PNe are descendants of stars that were formed up to a few Gyrs ago (Peimbert 1990, Maciel \& K\"oppen 1994). Another point is that, in the case of a high-metallicity medium, abundances derived in \hii\ regions can be biased due to significant temperature gradients in their interiors, while those derived in PNe are not (Stasi\'nska 2005). 
On the other hand, the chemical composition of PNe is not necessarily identical to that of the material out of which the progenitors were formed. It has long been known that, for example, the nitrogen observed in PNe has been partly produced   at various stages of the progenitor evolution (Renzini \& Voli 1981). Other elements may be affected as well, in smaller proportion but in a way that is far less well understood. Therefore, the comparison of abundances in PNe and \hii\ regions can also provide important clues on the nucleosynthesis in low- and intermediate-mass stars of different metallicities. 

In the case of spiral galaxies, \hii\ regions have revealed the presence of radial abundance gradients (Searle 1971, Shields 1974, McCall et al. 1985), which give clues on the chemical evolution of those galaxies. One of the important  questions to sort out is how these gradients evolve with time. Do they steepen? Do they flatten? Previous studies based on PNe in the Milky Way have led to contradictory results (Maciel \& K\"oppen 1994; Maciel et al. 2003; Stanghellini \& Haywood 2010). One difficulty in the study of abundance gradients from PNe located in the Milky Way is that their distances are in general very poorly known. This is why turning to other galaxies may be rewarding, despite the increased distance.

The face-on spiral galaxy  M33, at a distance of 840 kpc, has been thoroughly studied from the point of view of its ISM chemical abundance, both in terms of its \hii\ regions (Vilchez et al. 1988; Rosolowsky \& Simon 2008; Magrini et al. 2007, 2010; Bresolin 2011) and, more recently, its PNe (Magrini et al. 2009 for the entire range of galatic radii and Bresolin et al. 2010 for the inner zones). According to Magrini et al. (2010), PNe and \hii\ regions appear to show a similar metallicity gradient. 

NGC 300 is the next nearest, nearly face-on spiral galaxy, at a distance of 1.88 Mpc (Gieren et al. 2005). It is an Scd galaxy, quite similar to M33 in many respects, located in the Sculptor group. Its
main properties are listed in Table 1. Its giant \hii\ regions have been recently studied by Bresolin et al. (2009), using the FORS 2 spectrograph at the Very Large Telescope. These authors were able to determine  for the first time in this galaxy a radial gas-phase oxygen abundance gradient using only temperature-based abundance determinations, 
$-$0.077 dex/kpc, with a central abundance 12\,+\,log O/H $\simeq$
8.57. This agrees very well with the trend in metallicity obtained
for  A and  B supergiant stars  by  Urbaneja et al.
(2005) and Kudritzki  et al. (2008). In this paper, we use the same instrument to investigate the chemical composition of the brightest PNe and compact \hii\ regions in this galaxy. 
 In a separate paper, the analysis
of the pre-imaging (on-band and off-band \Oiii\ observations)
obtained with the same telescope  is presented (Pe\~na et al. 2012), and a list of candidate PNe is given. 

This paper is organized as follows. Section 2 presents the observational material and the data reduction. Section 3 presents the determination of the physical conditions, and the ionic and total abundances of our targets. Before getting to the heart of this paper, Section 4 revisits the classification of the candidate PNe and \hii\ regions. Section 5 presents classical abundance ratio diagrams for our NGC 300 objects. Section 6 presents the radial behavior of the abundance ratios derived from PNe and \hii\ regions. Section 7 provides our interpretation of the results, and Section 8 summarizes the main findings of our work.

\begin{table}[t!]
\begin{center}
\caption{Main properties of NGC\,300}
\label{log}
\begin{tabular}{lc}
\hline \hline
parameter & value \\
\hline
RA (J2000) & 00h 54m 53.48s \\
Dec (I2000) & $-$37$^o$41$'$03.8$''$ \\
Distance & 1.88 Mpc$^{a}$\\
$R_{25}$ & 9.75$'$ (5.33 kpc)$^{b}$ \\
Inclination & 39.8$^o$\\
Heliocentric radial velocity & 144\,km s$^{-1}$~$^{c}$ \\
$M^0_B$ & $-$17.88 \\
\hline
\multicolumn{2}{l}{$a$  Gieren et al. (2005)}\\
\multicolumn{2}{l}{$b$  HYPERLEDA database, Paturel et al. (2003)}\\
\multicolumn{2}{l}{$c$ de Vaucouleurs et al. (1991). } \\
\end{tabular}
\end{center}
\end{table}


\section{Observations and data reduction}
\label{sec:obs}

\subsection{Direct imaging}
In 2006 we started an observational program aimed at finding a large
number of PNe in NGC\,300 and analyzing their chemical properties.  The
Very Large Telescope (VLT) of the European Southern Observatory (ESO)
at Cerro Paranal,  equipped with the FORS\,2 spectrograph, was used for
this purpose. \Oiii\ on- and off-band pre-imaging data, was obtained using the
 FILT-500-5 and OIII/6000 filters, respectively, to search for emission
line objects. 
We covered two regions with the 6.8 $\times$ 6.8 arcmin$^2$ field of view of
FORS\,2: one, centered at
RA\,=\,00h 54m 49.0s, Dec\,=\,$-$37$^o$41$'$02.0$''$,  includes the center
of the galaxy, and the other, centered at RA\,=\,00h 55m 22.0s, Dec\,=\,$-$37$^o$43$'$00.0$''$, samples the galaxy outskirts (see Fig. 1 of Pe\~na et al. 2012).

After subtracting the on-band and off-band \Oiii\ images, we searched for
emission line objects in the two fields. Hundreds of such objects were
found. We considered as PN candidates  those objects that appear stellar with 
no stellar residual in the off-band
image. With such a criterion, we found more than 100 PN candidates: 
about 68 in the central field and about
34 in the external one.  We rediscovered most of the 34 PN candidates
reported by Soffner et al.  (1996).   The analysis
of the results obtained from our pre-imaging data is presented in Pe\~na et
al. (2012).
We selected about 50 objects (both PN candidates and
other emission line objects) for follow-up spectroscopy with FORS\,2
in multiobject (MXU) mode.

\begin{table} [h!]
\caption{Summary of spectroscopic observations$^{a}$}
\begin{tabular}{cccrr}
\hline \hline
field & grism & filter& exp. time \\
\hline
center    & 600B+22 & none& 7$\times$1800 s\\
center  & 600RI+19 & GG435 & 6$\times$1200\,s + 1219\,s \\
center & 300I+21 & OG590 & 4$\times$1800\,s\\
outskirts   & 600B+22 & none & 6$\times$1800\,s + 1200\,s \\
outskirts & 600RI+19 & GG435 & 3$\times$1800\,s \\
outskirts & 300I +21& OG590 & 1800\,s\\
\hline
\multicolumn{4}{l}{$a$ Observing dates: August 19 - 22, 2006}
\end{tabular}
\end{table}

\subsection{Spectroscopy}
 Multiobject  spectra  were obtained on August 19-22, 2006.  Masks with 1 arcsec-wide slit were designed for the FORS2 MXU mode to observe 
as many targets as possible without affecting the desirable wavelength range,
in the two regions of the galaxy covered by our narrow-band images.  Of course 
the brightest PN candidates were chosen first, but a certain number of faint objects 
were included too. The range of log F(5007) values for the observed PNe covers
from $-$14.693 to $-$16.485  which correspond to  \Oiii\ magnitudes $m$(5007) from -23  to -27.5.
 Three grisms were used in order to
cover the widest wavelength range possible: 600B (approximate range
3600-5100 \AA,  4.5 \AA~FWHM spectral  resolution), 600RI
(approximate range 5000-7500 \AA,  5 \AA~ spectral resolution),  and 300I
(approximate range 6500-9500 \AA, 10 \AA~spectral resolution). Multiple
exposures were carried out for each setup. Our 
exposures are listed in Table 2.  Total exposure times for the central
field  were 3.5 h for grism 600B,  3.3 h for grism 600RI and 2 h
for grism 300I. For the external field the exposure times were
3.33 h for grism 600B, 1.5 h for grism 600RI and 0.5 h for grism
300I.
Three spectroscopic standard stars (EG\,274, LDS\,749B and BMP\,16274)
were observed each night, through a 5$''$-wide slit, for flux
calibration.   Seeing conditions were in general very good (0.7$''$ -
0.9$''$), except for the first night (August 19-20), when the seeing
was larger than 1$''$.  Frames containing the same spectral
data were combined and then reduced with  standard procedures by
using the EsoRex pipeline provided by ESO: bias subtraction, flat
fielding, and wavelength calibration were applied at this stage.
Afterwards we used standard IRAF\footnote{IRAF is distributed by the
NOAO, which is operated by the AURA, Inc., under cooperative agreement
with the NSF.}  tasks for spectral extraction and flux calibration.

In Fig.\ref{fig:spectra}  the  calibrated spectra (showing the full wavelength range)
for all the observed objects  are shown.

\subsubsection{Measuring and dereddening the fluxes}

Line intensities were measured in the flux-calibrated spectra with the task 
{\it splot} in IRAF by integrating between two positions over a continuum level estimated 
by eye (actually there is no detectable continuum in most of the objects, in particular PNe). 
Then we proceeded to calculate the logarithmic reddening correction, c(H$\beta$) from the 
Balmer decrement. Due to the fact that we have three independent  spectral ranges (blue - 600B, 
visible - 600RI and red - 300I) that can be affected by differences in flux calibration, we determined the reddening from the blue spectra, where   
H$\beta$, H$\gamma$, H$\delta$ and the upper lines of the Balmer series are present.    
The logarithmic reddening correction, c(H$\beta$),  was computed from these lines  by assuming 
case B recombination theory (Storey \& Hummer 1995). The  lines measured in the blue spectral 
range were dereddened with this c(H$\beta$) value and by adopting the Seaton (1979) reddening law. 
In the 600RI and 300I spectra only H$\alpha$ is available among the H Balmer series lines, so we  adopted
 the c(H$\beta$) value obtained in the blue to deredden the lines in these spectral range, 
 relative to H$\alpha$ (again Seaton's reddening law was used).  Afterwards we matched the 
 blue, visible and red portions of the spectra by assuming a dereddened H$\alpha$/H$\beta$ 
 flux ratio as given by recombination theory [I(H$\alpha$)/I(H$\beta$) $\sim$ 2.86].
 The line flux uncertainties  (as well as upper limits) were determined as $2\times$  
the rms noise measured  on each side of the line. 

In Table \ref{tab:intensities} we present the   dereddened line fluxes and associated uncertainties for all the
lines detected in each object. 
For  lines that are important
abundance diagnostics upper limits are given.
At the bottom of the table we also list the c(H$\beta$) values  and the logarithm of the total
observed H$\beta$ flux (since the full table is available in the on-line version only, here we show 
just an example, for illustration purposes). The deprojected galactocentric distance of each object,
 given in terms of the 25th magnitude B-band isophotal radius $R_{25}$ = 9.75$'$, and  
 calculated by adopting the parameters given in Table 1, is also included in Table 3. 
 
It is interesting to notice that c(H$\beta$) is very low in all our targets, in particular in PNe 
where almost no reddening was found. Only a few compact \hii\ regions show significant reddening. 
This is consistent with the low foreground E(B$-$V) values determined by other authors,
ranging between  0.013 (Schlegel et al. 1998) and  0.096 
(Gieren et al. 2005).

\begin{table*} [t!]
\caption{Line intensities  for \hii\ regions and PNe, relative to H$\beta$=1.00$^{1,2,3}$. Full table only available at the CDS}
\label{tab:intensities}
\begin{center}
\begin{tabular}{lrrrrrrcccccccc}
\hline \hline
\multicolumn{2}{l}{Object number$^2$}&\multicolumn{2}{c}{12}&\multicolumn{2}{c}{14}&\multicolumn{2}{c}{20}\\
\multicolumn{2}{l}{RA (h m s)}&\multicolumn{2}{c}{00 54 37.89}&\multicolumn{2}{c}{00 54 38.91}&\multicolumn{2}{c}{00 54 41.59}\\
\multicolumn{2}{l}{Dec ($^o$ $'$ $''$) }&\multicolumn{2}{c}{$-$37 40 14.0}&\multicolumn{2}{c}{$-$37 39 43.2}&\multicolumn{2}{c}{$-$37 40 21.4}\\
\multicolumn{2}{l}{$R/R_{25}$}&\multicolumn{2}{c}{0.33}&\multicolumn{2}{c}{0.33}&\multicolumn{2}{c}{0.25}\\
\multicolumn{2}{l}{Object type} &\multicolumn{2}{c}{PN}&\multicolumn{2}{c}{PN}&\multicolumn{2}{c}{PN}\\
\noalign{\smallskip} \noalign{\hrule} \noalign{\smallskip}
$\lambda$ &ion&I/I(Hb)&error&I/I(Hb)&error)&I/I(Hb)&error&\\
\hline
3727+&[OII]&0.608&0.150&0.643&0.050&$<$0.870&\\
3833.4&H9&&&0.078&0.039&&&\\
3868.7&[NeIII]&0.674&0.168&0.796&0.040&&&\\
3889.1&HeI+H8& &&0.165&0.033&&&\\
3970.5&[NeIII]+H7& &&0.450&0.045&&&\\
4026.2&HeI& &&0.043&0.022&&&\\
4068.6&[SII]&$<$0.10&&0.061&0.031&&&\\
4101.7&Hd&0.196&0.098&0.256&0.015&0.255&0.033\\
4267.1&CII&$<$0.10&&$<$0.050&&&&\\
4340.5&Hg&0.486&0.050&0.478&0.020&0.474&0.033\\
4363.2&[OIII]&0.172&0.060&0.212&0.020&$<$0.075&\\
4471.5&HeI&    &&0.044&0.022&&&\\
4686.7&HeII&0.632&0.095&0.245&0.015&$<$0.027&\\
4711.4&[ArIV]&$<$0.10&&$<$0.030&&&&\\
4740.2&ArIV&$<$0.10&&0.078&0.039&&&\\
4861.3&Hb&1.000&0.060&1.000&0.030&1.000&0.040\\
4958.9&[OIII]&4.700&0.280&3.325&0.033&2.314&0.070\\
5006.8&[OIII]&14.38&0.860&9.690&0.097&6.772&0.20\\
5754.7&[NII]&&&&&&&&\\
5875.6&HeI&&&0.148&0.013&0.086&0.009\\
6300.3&[OI]&&&0.057&0.008&&&\\
6312.1&[SIII]&&&0.022&0.005&&&\\
6548.3&[NII]&&&0.174&0.015&0.104&0.010\\
6562.8&Ha&2.860&0.080&2.855&0.020&2.860&0.080\\
6583.4&[NII]&0.160&0.080&0.526&0.010&0.283&0.023\\
6678.1&HeI&&&0.033&0.006&0.035&0.007\\
6716.5&[SII]&&&0.041&0.012&0.118&0.010\\
6730.8&[SII]&&&0.085&0.007&0.120&0.010\\
7065.3&HeI&&&0.049&0.009&0.106&0.090\\
7135.8&[ArIII]&&&0.151&0.008&0.186&0.015\\
7319+&[OII]&&&0.100&0.010&0.042&0.007\\
7330+&[OII]&&&0.055&0.020&0.077&0.011\\
9068.6&[SIII]&&&0.350&0.070&0.533&0.060\\
9529.8&[SIII]&&&0.785&0.150&1.207&0.120\\
\hline
\multicolumn{2}{l}{c(Hb)}&0.00&&0.00&&1.17&\\
\multicolumn{2}{l}{log F(Hb)$^3$}&$-$16.590&& $-$15.920&&$-$16.350&\\
\hline
\multicolumn{8}{l} {$^{1}$ $<$: upper limits, +: both lines, : very uncertain}\\
\multicolumn{8}{l} {$^{2}$ \# as in Table 2 of Pe\~na et al. 2012}\\
\multicolumn{8}{l} {$^{3}$  units of erg cm$^{-2}$ s$^{-1}$}\\
\hline
\end{tabular}
\end{center}
\end{table*}


\begin{figure*}
\includegraphics[width=0.9\textwidth]{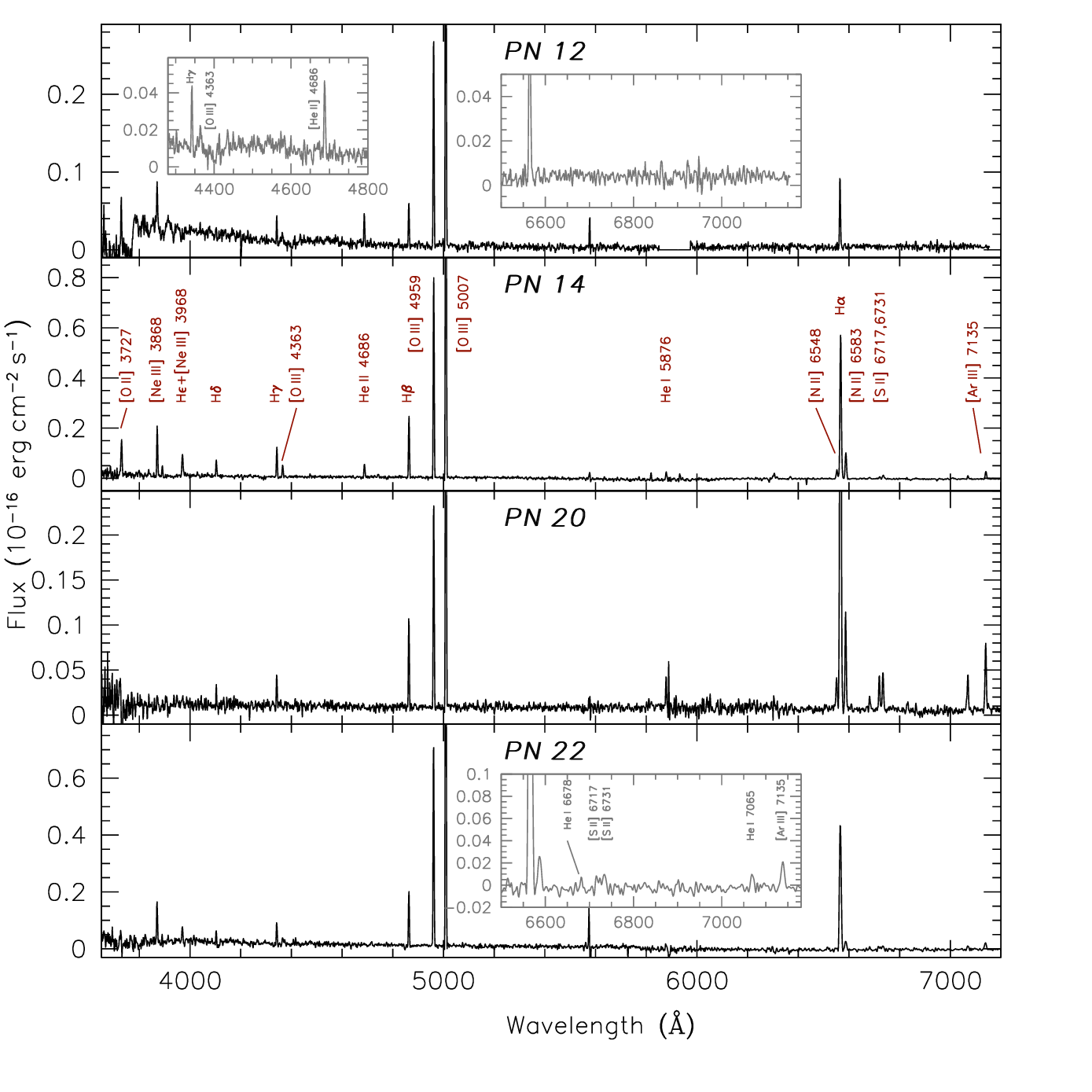} 
\caption{Calibrated spectra of our observed PNe and compact \hii\ regions (Figure  available electronically only). \label{fig:spectra}}
\end{figure*}


\setcounter{figure}{0}
\begin{figure*}
\includegraphics[width=0.9\textwidth]{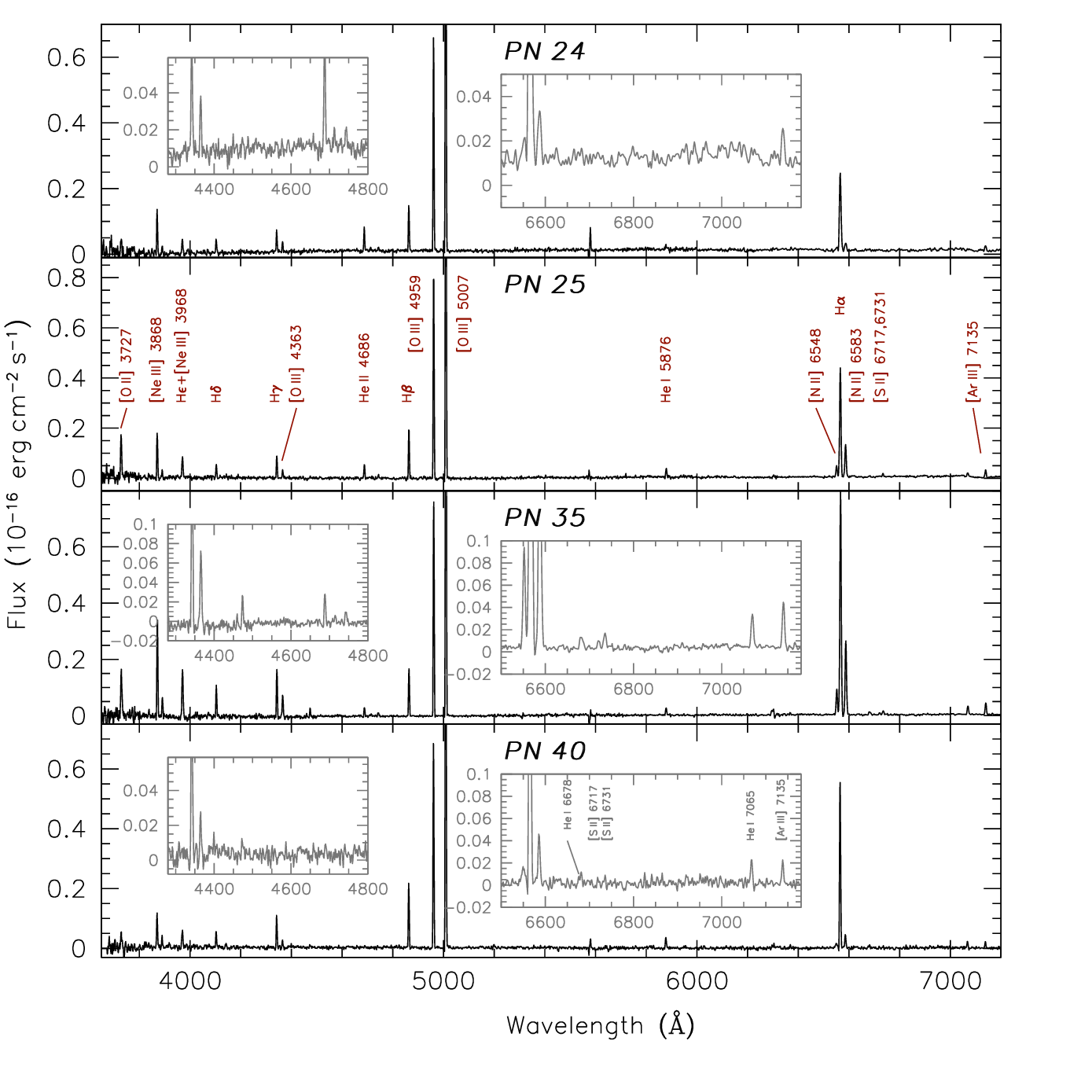} 
\caption{ \it (continued). }
\end{figure*}


\setcounter{figure}{0}
\begin{figure*}
\includegraphics[width=0.9\textwidth]{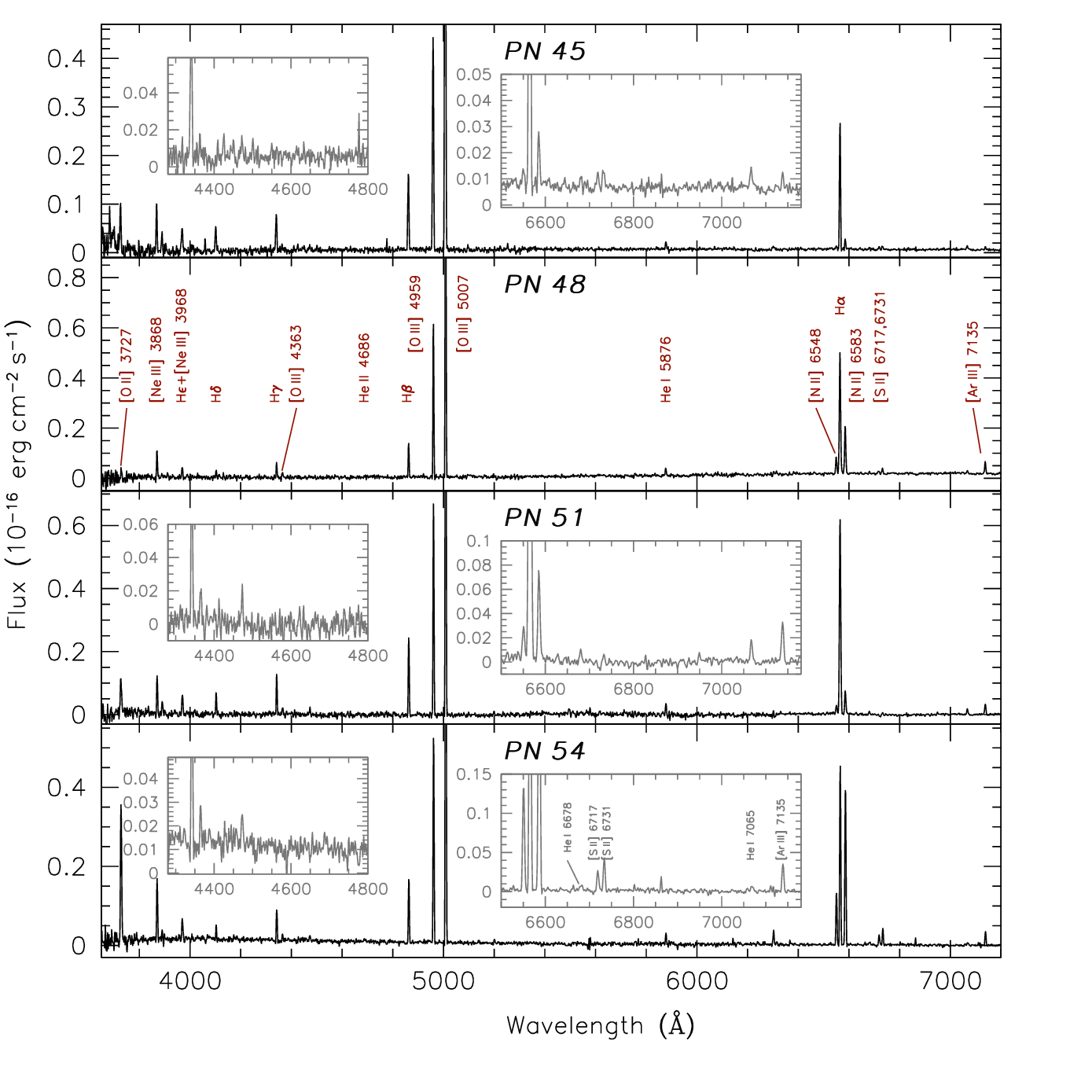} 
\caption{ \it (continued). }
\end{figure*}


\setcounter{figure}{0}
\begin{figure*}
\includegraphics[width=0.9\textwidth]{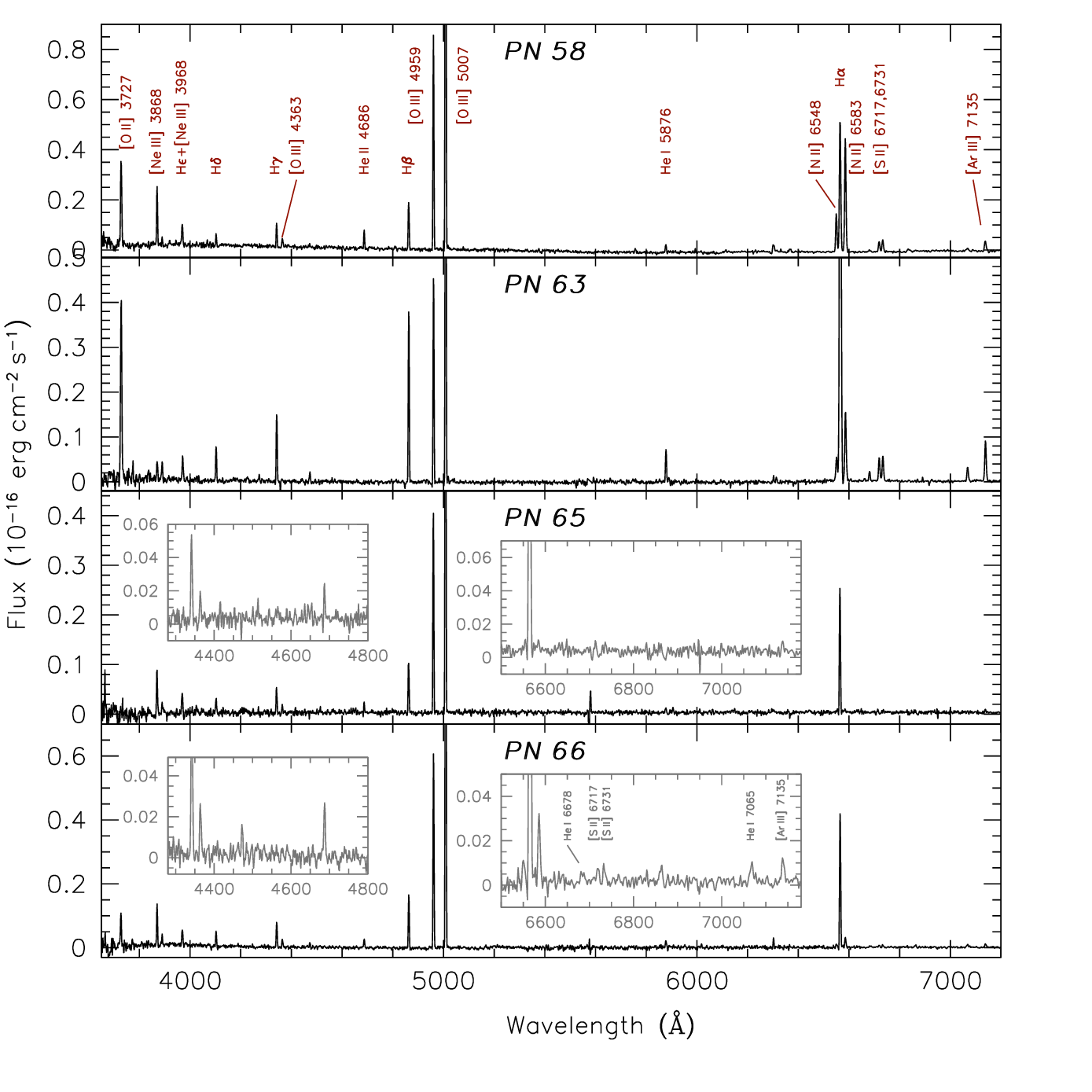} 
\caption{ \it (continued). }
\end{figure*}


\setcounter{figure}{0}
\begin{figure*}
\includegraphics[width=0.9\textwidth]{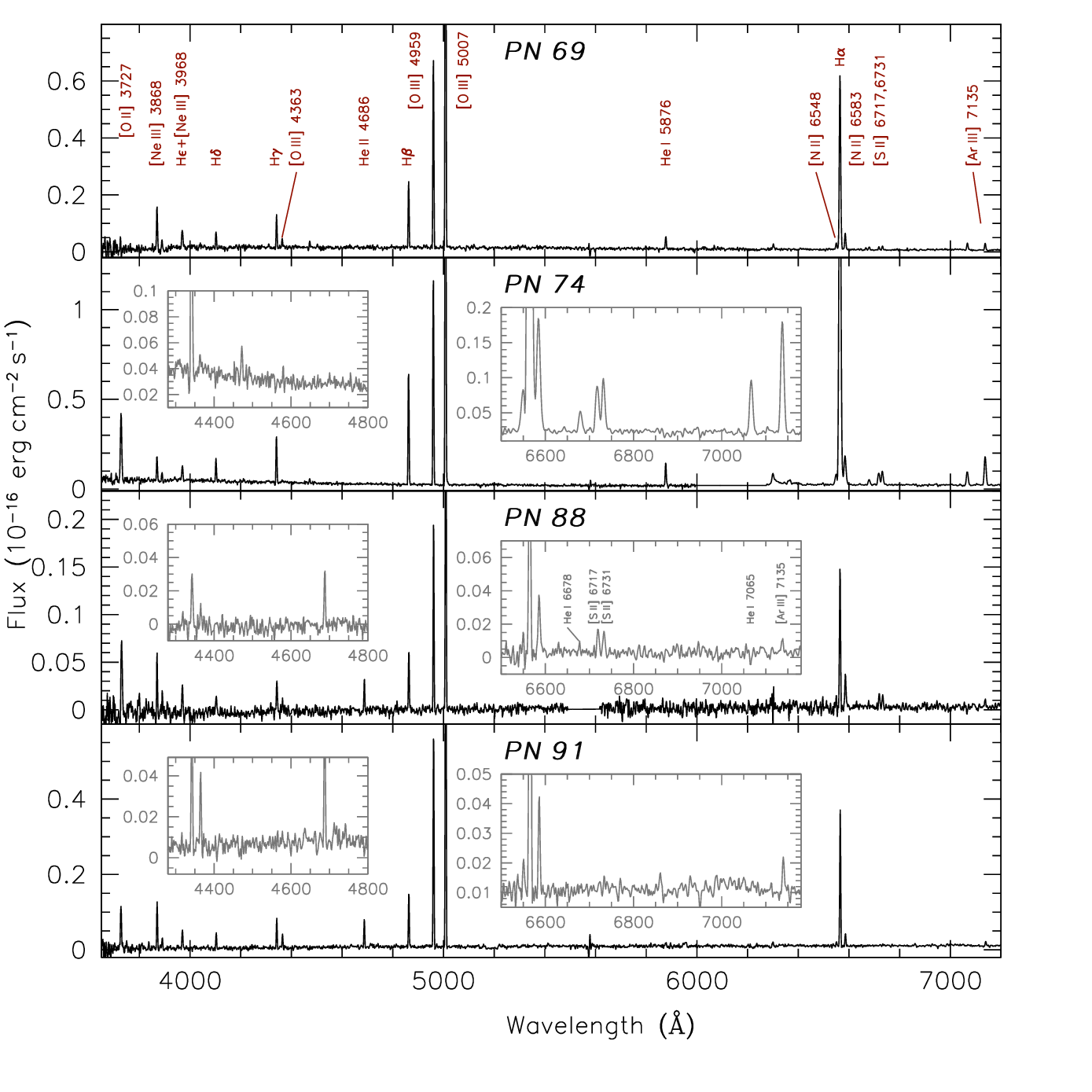} 
\caption{ \it (continued).  \rm Note that for PN 74, the red spectrum has been obtained with grism 300I, which has a much lower resolution than 600B and 600RI. }
\end{figure*}


\setcounter{figure}{0}
\begin{figure*}
\includegraphics[width=0.9\textwidth]{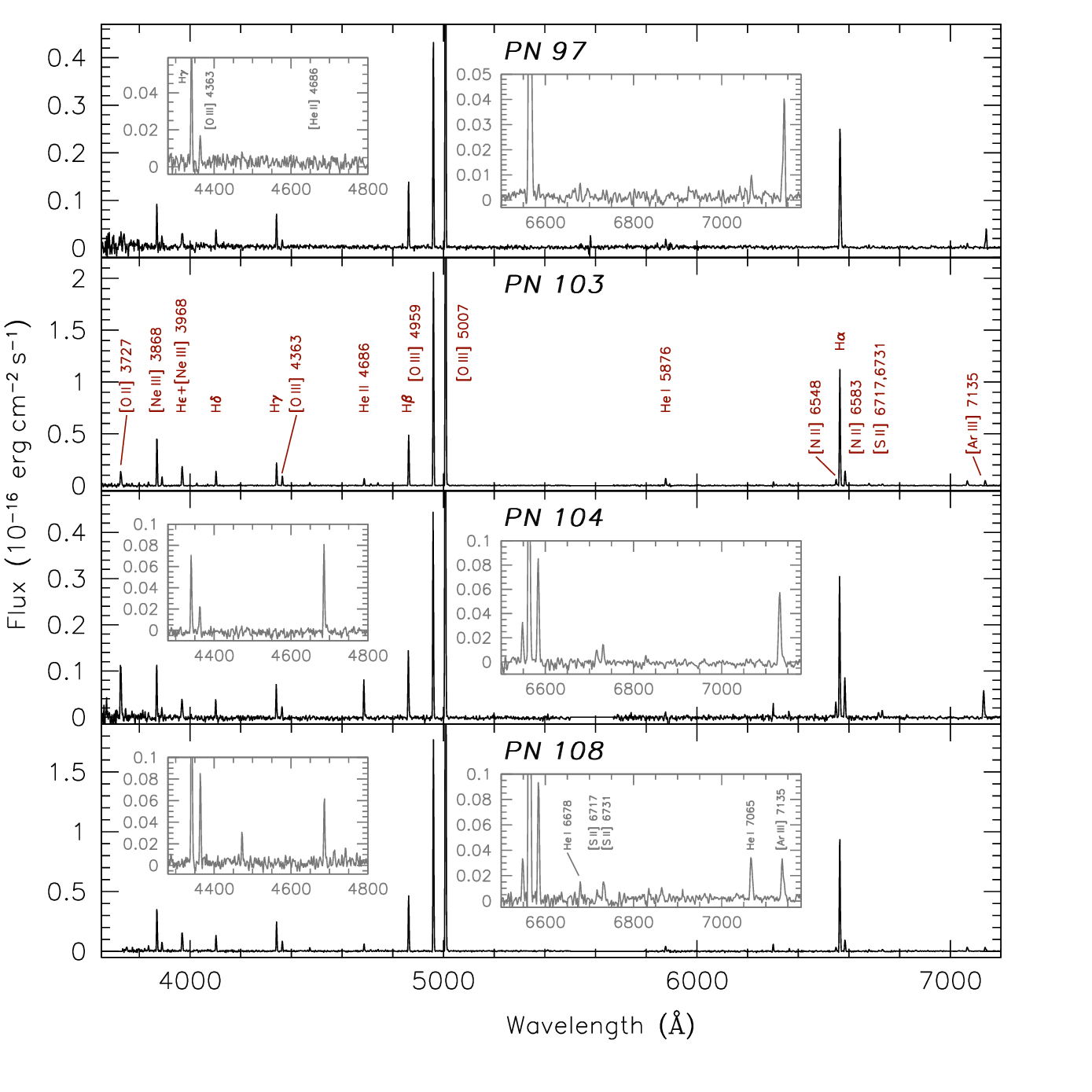} 
\caption{ \it (continued). }
\end{figure*}


\setcounter{figure}{0}
\begin{figure*}
\includegraphics[width=0.9\textwidth]{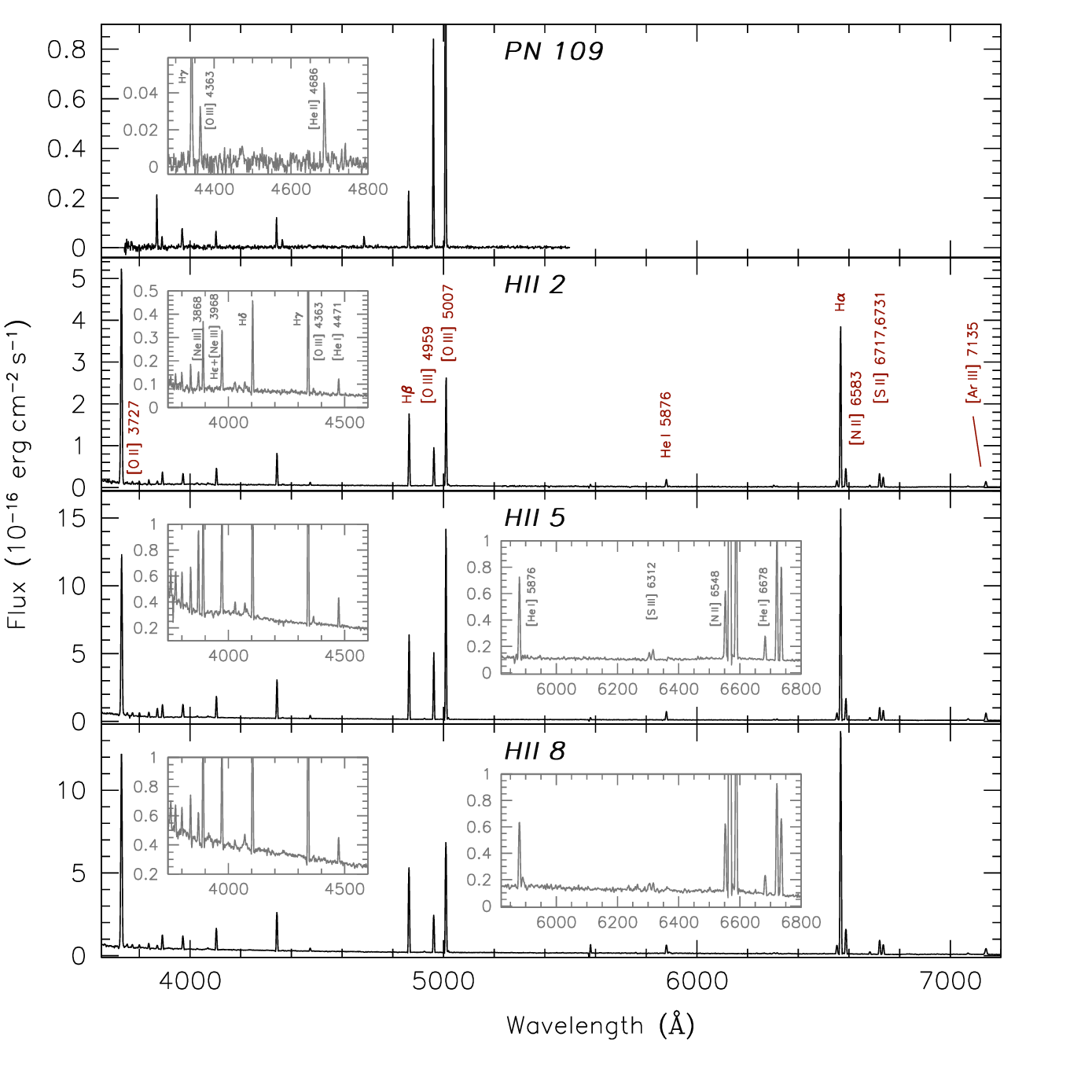} 
\caption{ \it (continued). }
\end{figure*}


\setcounter{figure}{0}
\begin{figure*}
\includegraphics[width=0.9\textwidth]{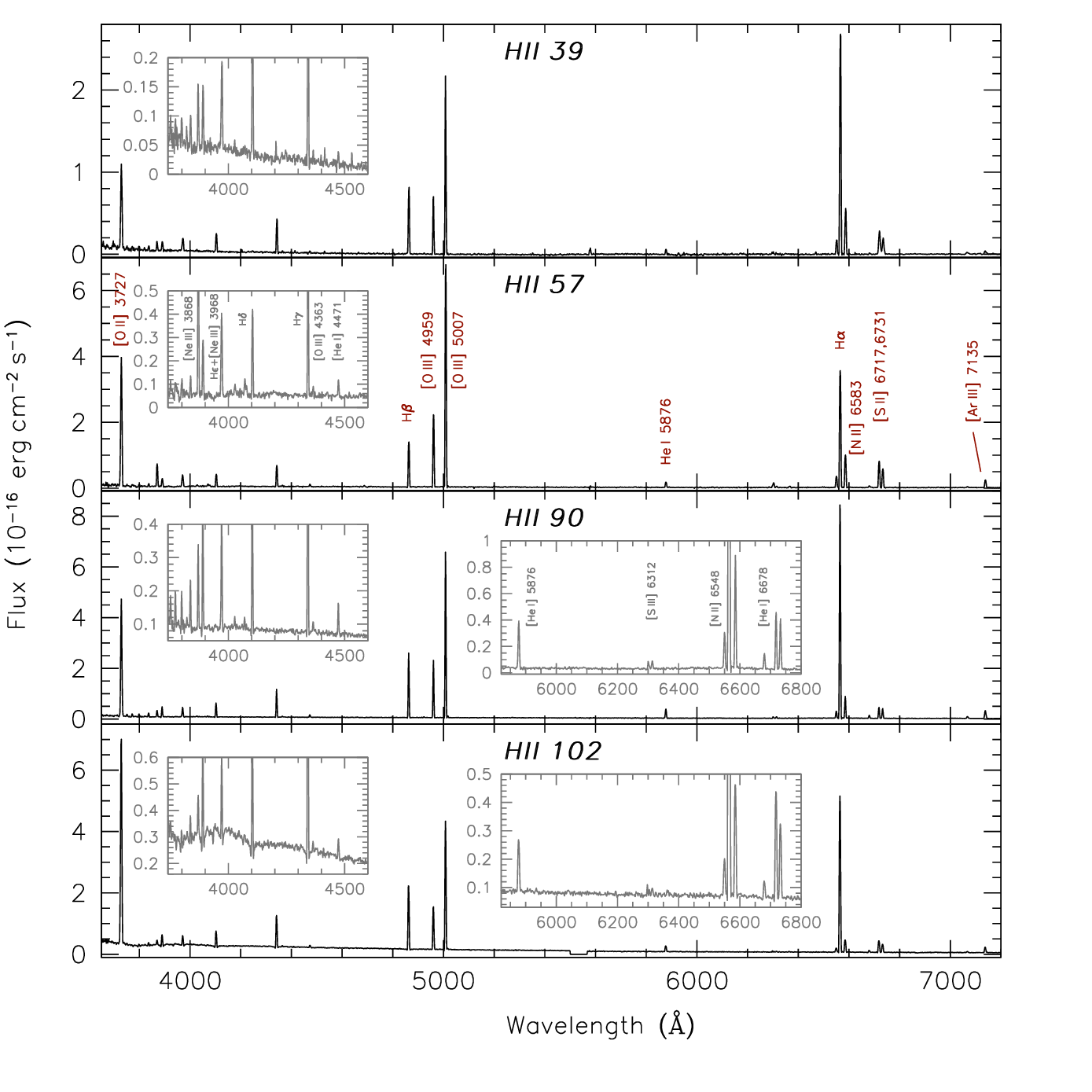} 
\caption{ \it (continued). }
\end{figure*}



%

%

\section {Determination of physical conditions, ionic and total abundances}
\label{sec:phiab}

\subsection{Atomic data}
\label{sec:atomdat}

\begin{table*} [h!]
\caption{Atomic data used in our abundance determinations}
\label{tab:atomdat}
\begin{tabular}{lll}
\hline \hline 
ion & collision strengths & radiative transition probabilities\\
\hline 
O\,{\sc ii} & Kisielius et al. (2009) & Zeippen (1982) \\
O\,{\sc iii} & Aggarwal \& Keenan (1999) & Galavis et al. (1997),  Storey \& Zeippen (2000) \\
N\,{\sc ii} & Tayal (2011) &   Galavis et al. (1997) \\
Ne\,{\sc iii} &  McLaughlin \& Bell 2000 & Galavis et al. (1997) \\
S\,{\sc ii} & Tayal \& Zatsarinny (2010) & Mendoza \& Zeippen (1982) \\
S\,{\sc iii} & Tayal \& Gupta (1999) & Froese Fischer et al. (2006) \\
Ar\,{\sc iii} &  Munoz Burgos et al. (2009)  &    Munoz Burgos et al. (2009) \\
Ar\,{\sc iv} & Ramsbottom et al. (1997)  &   Mendoza \& Zeippen (1982)  \\
\hline
\end{tabular}
\end{table*}

Since the atomic data adopted in the computation of abundances translate immediately into the results, we present them first. A number of new atomic calculations has  recently become available, attempting to improve on previous results. One emblematic case is that of \Op, which is used as a density diagnostic, and for which  the collision strenghts computed by McLaughlin \& Bell (1998)  gave results inconsistent with those of other ions, as shown by Copetti \& Writzl (2002) and Wang et al. (2004). This prompted new atomic computations and a thorough discussion to address this problem (Kisielius et al. 2009). The latter authors also recommend the use of the transition probabilities from Zeippen (1982).

\begin{figure}
\includegraphics[width=6 cm]{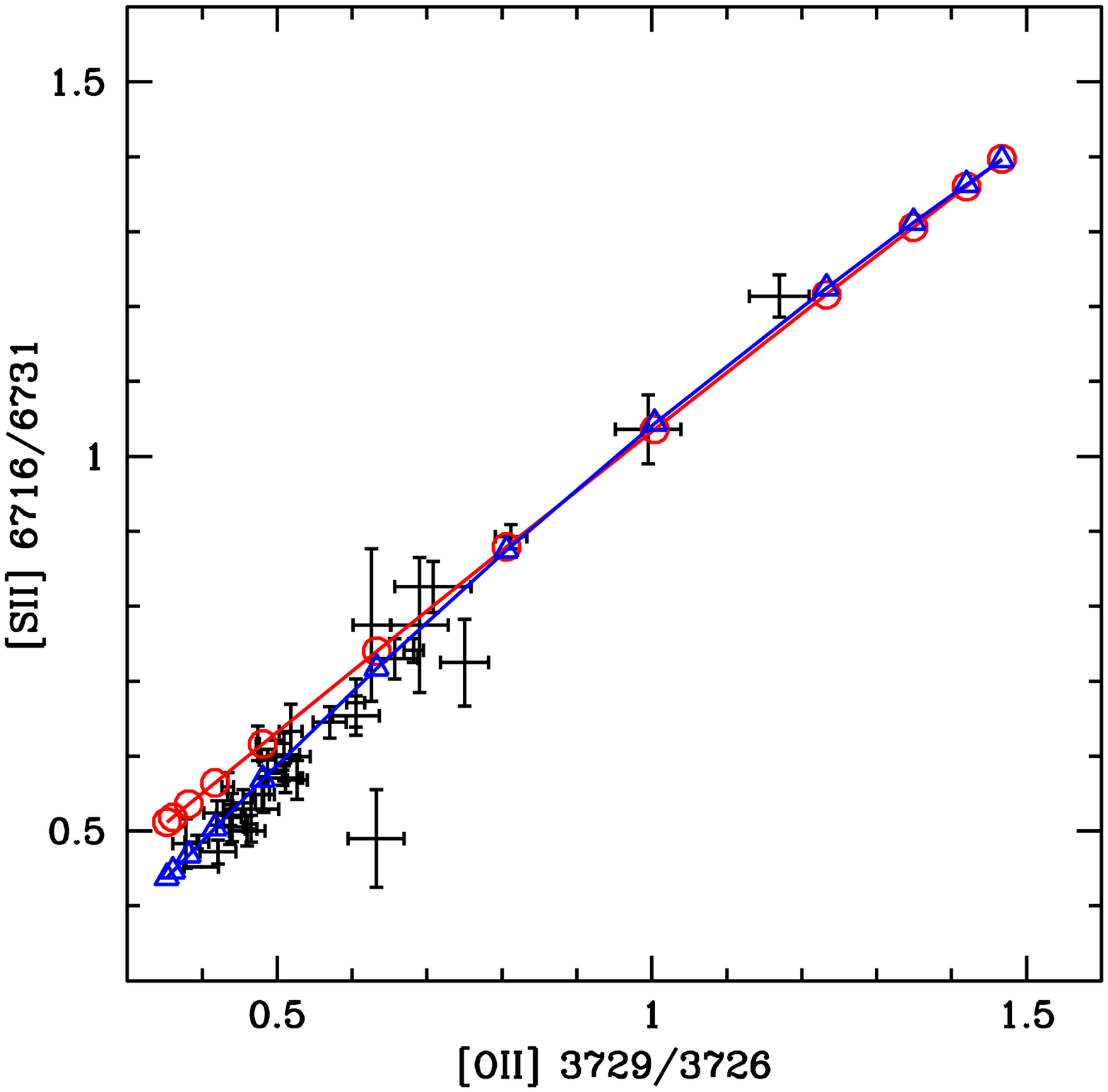} 
\caption{Values of the density indicator \rSii\ as a function of \rOii. Points with error bars: observed values in the sample of PNe by Wang et al. (2004).  Continuous lines: lines of equal electron density at a temperature of $10^4$\,K from $n_{\rm e}=20$\cmcub\ (upper right) to $n_{\rm e}=10^5$\cmcub\ (lower left). Red: using the \sii\ transition  probabilities  by Tayal \& Zatsarinny (2010). Blue:  using the \sii\ transition  probabilities  from Mendoza \& Zeippen (1982).\label{fig:siioii}}
\end{figure}

New collision strenghts and transition probabilities  have  been computed for \Sp\ by Tayal \& Zatsarinny (2010). We used the  \rSii\ and \rOii\ line ratios observed by Wang et al. (2004) in Galactic PNe to check the densities obtained using the Tayal \& Zatsarinny (2010) atomic data. In Fig. \ref{fig:siioii} we show the data with their error bars in the \rSii\ vs. \rOii\ plane, together with the line of equal electron density at a temperature of $10^4$\,K for densities ranging from $n_{\rm e}=20$\cmcub\ (upper right) to $n_{\rm e}=10^5$\cmcub\ (lower left), shown in red. It is clear that, at the high-density end -- the most drastic regime in our study, as will be seen later -- the observational points lie well below this line, meaning that the  \rSii\ ratio indicates a subsantially higher density than the \rOii\ one (by factors as large as 10), and many of the observed values of \rSii\ are outside the high density limit. Even if one does not expect the mean densities in the \Op\ and in the \Sp\ region to be identical, due to the fact that a density gradient is likely to exist and that the \Op\ and \Sp\ zones are not exactly coextensive, such a large difference looks unrealistic. We have therefore taken the \sii\  transition  probabilities  from Mendoza \& Zeippen (1982), which lead to the line of equal density shown in blue. This line goes beautifully through the observational points, and therefore we adopted this reference for the transition probabilities of \sii.

New atomic data are also available for \Np\ and \Arpp, which we use in the present work. McLaughlin et al. (2011) announced new collision strenghts for \Nepp\ but the data are not available yet. 
We list in Table \ref{tab:atomdat} the references for the atomic data used in the present work for ions of O, N, Ne, S and Ar. Note that for \Op\ we preferred the transition probabilities from Zeippen (1982) over more recent estimates,  as recommended by Kisielius et al. (2009).

\begin{figure}
\includegraphics[width=10cm,trim=30mm 80mm 0mm 70mm, clip=true ]{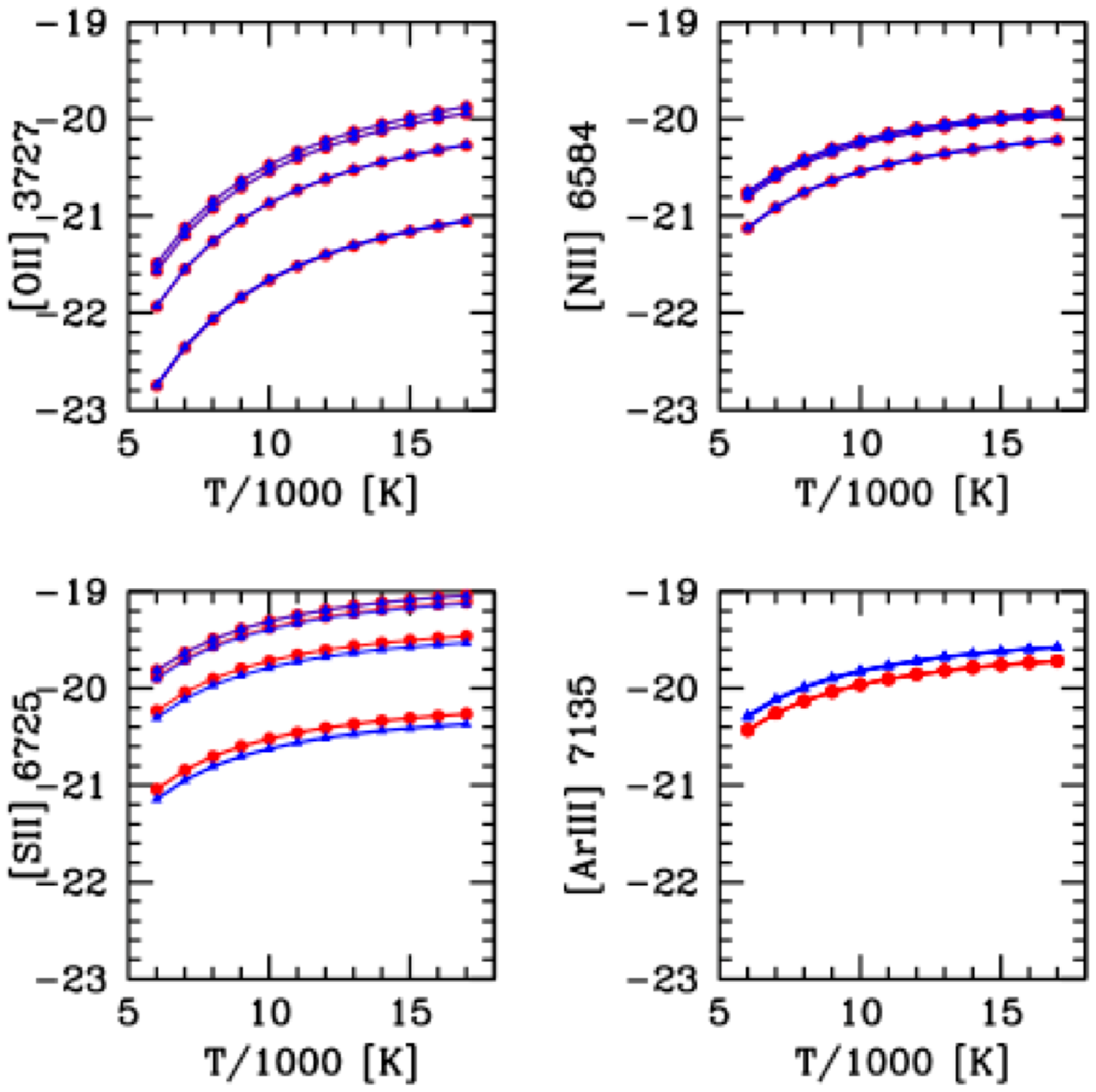} 
\caption{Logarithms of the emissivities of the \Oii, \Nii, \Sii, and \Ariii\ lines  for densities $n_{\rm e}\,=\,10^2, 10^3, 10^4$ and $10^5$\cmcub. Red: with the atomic data used in the present work. Blue:  with the atomic data used in our previous work. \label{fig:emissivities}}
\end{figure}

The emissivities of the different lines were obtained with a 5-level atom in each case. In Fig. \ref{fig:emissivities} we compare the present emissivities of the \Oii, \Nii, \Sii, and \Ariii\ lines in nebular conditions (red curves), with respect to those obtained with  atomic data used in our previous work (Pe\~na et al. 2007, Bresolin et al. 2009, 2010) (blue curves), for densities $n_{\rm e}=10^2, 10^3, 10^4$ and $10^5$\cmcub\ (the lower curves corresponding to lower emissivities).  Overall, the differences are not very large, the largest ones reaching 0.2 dex for \ariii. Incidentally, this plot also shows how much the \Oii\ emissivity depends on density as soon as it exceeds $10^3$\cmcub, meaning that  it is essential to have a good determination of the density for a proper determination of O/H and  N/O.

For \Hp\ and \Hepp\  we use the case B emissivities from Hummer \& Storey (1987), and for \Hep\ we use the emissivities from Porter et al. (2007), including collisional excitation. In the case of optically thin PNe, case A would be more appropriate, implying i) a possible error in the helium abundance and ii) a possible underestimate of heavy element abundances with  respect to hydrogen by up to 30\%. However we do not expect such objects to be present in our sample, which is composed of  luminous PNe, which are likely to be rather optically thick (see Stasi\'nska et al. 1998). Many of our objects show a detectable \oi\ line giving observational support to such a claim (although the possibility of density-boundedness in some directions cannot be discarded). 

\subsection{Plasma diagnostics and ionic abundances}
\label{sec:plasmadiag}

In Table \ref{tab:plasma} we list the values of the electron temperatures derived from the \rOiii\ and \rNii line ratios obtained from Table \ref{tab:intensities}, as well as the densities derived from \rSii\ and \rAriv. We give in brackets the 15 and 85 percentiles, obtained by a Monte Carlo procedure as explained in Sect. \ref{sec:uncertainties}. This means that there is a 70\% probability that the true value lies between those percentiles (to be compared to the 68.3\% probability to be within one standard deviation for a Gaussian distribution). Similarly to Table \ref{tab:intensities},  Table \ref{tab:plasma} is ordered by object number, irrespective of the nature of the object (PN or compact \hii\ region). The first column indicates the object type as assigned in Sect. \ref{sec:distinguish}: PN for planetary nebulae and H\,{\sc ii} for compact \hii\ regions.  


\begin{table*} [h!]
\caption{Electron densities and  temperatures for the PNe and compact \hii\ regions in   NGC\,300 (table  available electronically only).}   
\label{tab:plasma}\centering\begin{tabular}{lcccc}
\hline \hline 
\smallskip
 object   &   $n[$S\,{\sc ii}]$^a$ &  $n$[Ar\,{\sc iv}]$^a$ &     $T$[O\,{\sc iii}]$^a$  &   $T$[N\,{\sc ii}]$^a$   \\
          &   (cm$^{-3}$) &  (cm$^{-3}$) &   (K)  &   (K)    \\
\hline
H\,{\sc ii}  2         &  3.9$\times 10^{1}$ (3.0$\times 10^{1}$, 6.5$\times 10^{1}$) &        ---     & 11458  ( 9647, 12768) &            ---            \\
H\,{\sc ii}  5         &  1.3$\times 10^{2}$ (7.1$\times 10^{1}$, 1.7$\times 10^{2}$) &             ---                &  9131  ( 8400, 9688) &            ---            \\
H\,{\sc ii}  8         &  3.0$\times 10^{1}$ (3.0$\times 10^{1}$, 5.4$\times 10^{1}$) &               ---              &  9791  ( 8177, 10930) &            ---            \\
PN  12         &               ---               &             ---                & 12249  (10726, 13847) &            ---            \\
PN  14         &  1.8$\times 10^{4}$ (4.9$\times 10^{3}$, 1.0$\times 10^{5}$) &            ---                 & 15446  (13858, 16035) &            ---            \\
PN  20         &  6.4$\times 10^{2}$ (4.4$\times 10^{2}$, 9.0$\times 10^{2}$) &               ---            &           ---             &            ---            \\
PN  22         &  4.0$\times 10^{2}$ (8.6$\times 10^{1}$, 9.7$\times 10^{2}$) &               ---            & 15895  (14362, 17040) &            ---            \\
PN  24         &              ---              & 9.5$\times 10^{3}$ (9.3$\times 10^{3}$, 9.7$\times 10^{3}$) & 13870  (12802, 14892) &            ---            \\
PN  25         &  4.2$\times 10^{3}$ (1.7$\times 10^{3}$, 1.0$\times 10^{5}$) & 2.2$\times 10^{4}$ (2.2$\times 10^{4}$, 2.3$\times 10^{4}$) & 12735  (11920, 13332) &            ---            \\
PN  35         &  6.2$\times 10^{3}$ (1.9$\times 10^{3}$, 1.0$\times 10^{5}$) & 1.6$\times 10^{4}$ (1.5$\times 10^{4}$, 1.7$\times 10^{4}$) & 14575  (13736, 15387) &            ---            \\
H\,{\sc ii} 39         &  3.0$\times 10^{1}$ (3.0$\times 10^{1}$, 1.3$\times 10^{2}$) &               ---            &  9486  ( 8516, 10401) &            ---            \\
PN  40         &              ---              &               ---            & 12345  (11273, 13434) &            ---            \\
PN  45         &  2.2$\times 10^{3}$ (1.3$\times 10^{3}$, 4.7$\times 10^{3}$) &               ---            &  9886  ( 8153, 11093) &            ---            \\
PN  48         &  2.0$\times 10^{3}$ (9.9$\times 10^{2}$,4.9$\times 10^{3}$) &               ---            & 13691  (12215, 14910) &            ---            \\
PN  51         &              ---              &               ---            & 12010  (11277, 12608) &            ---            \\
PN  54         &  3.3$\times 10^{3}$ (2.4$\times 10^{3}$, 4.4$\times 10^{3}$) &               ---            & 11791  (11288, 12231) & 12665  ( 9198, 16327) \\
H\,{\sc ii} 57         &  3.0$\times 10^{1}$ (3.0$\times 10^{1}$, 7.1$\times 10^{1}$) &               ---            &  9515  ( 8803, 10225) &            ---            \\
PN  58         &  1.2$\times 10^{3}$ (9.0$\times 10^{2}$, 1.7$\times 10^{3}$) &               ---            & 13648  (13121, 14135) & 12960  (11011, 15234) \\
PN  63         &  8.6$\times 10^{2}$ (6.5$\times 10^{2}$, 1.2$\times 10^{3}$) &               ---            & 10075  ( 8768, 11462) &            ---            \\
PN  65         &              ---              &               ---            & 12618  (12031, 13107) &            ---            \\
PN  66         &  1.3$\times 10^{2}$ (3.0$\times 10^{1}$, 5.9$\times 10^{2}$) &               ---            & 12696  (12302, 12999) &            ---            \\
PN  69         &  1.5$\times 10^{3}$ (8.5$\times 10^{2}$, 3.7$\times 10^{3}$) &               ---            & 12886  (11947, 13569) &            ---            \\
PN  74         &  6.3$\times 10^{2}$ (5.3$\times 10^{2}$, 7.7$\times 10^{2}$) &               ---            & 10147  ( 8393, 11276) &            ---            \\
H\,{\sc ii} 87         &  1.6$\times 10^{2}$ (1.2$\times 10^{2}$, 2.1$\times 10^{2}$) &               ---            & 11325  (10541, 12001) &            ---            \\
PN  88         &  6.2$\times 10^{2}$ (6.0$\times 10^{2}$, 6.5$\times 10^{2}$) &               ---            & 14191  (12963, 15422) &            ---            \\
H\,{\sc ii} 90a        &  2.8$\times 10^{2}$ (2.7$\times 10^{2}$, 2.8$\times 10^{2}$) &               ---            & 10551  ( 9461, 11305) &            ---            \\
H\,{\sc ii} 90b        &  1.2$\times 10^{3}$ (1.1$\times 10^{3}$, 1.2$\times 10^{3}$) &               ---            & 10317  ( 9270, 11138) &            ---            \\
PN  91         &  1.2$\times 10^{4}$ (1.1$\times 10^{4}$, 1.3$\times 10^{4}$) & 1.7$\times 10^{3}$ (1.6$\times 10^{3}$, 1.8$\times 10^{3}$) & 14906  (14419, 15342) &            ---            \\
PN  96         &              ---              &               ---            & 14318  (13640, 14873) &            ---            \\
PN  97         &              ---              &               ---            & 12010  (10572, 13505) &            ---            \\
H\,{\sc ii} 102        &  3.0$\times 10^{1}$ (3.0$\times 10^{1}$, 3.0$\times 10^{1}$) &               ---            & 11146  (10076, 12190) &            ---            \\
PN 103         &  1.0$\times 10^{5}$ (1.0$\times 10^{5}$, 1.0$\times 10^{5}$) & 1.2$\times 10^{4}$ (1.1$\times 10^{4}$, 1.3$\times 10^{4}$) & 13401  (12848, 13820) &            ---            \\
PN 104         &  1.5$\times 10^{3}$ (1.4$\times 10^{3}$, 1.5$\times 10^{3}$) &               ---            & 14969  (14106, 15725) &            ---            \\
PN 108         &  8.8$\times 10^{3}$ (8.6$\times 10^{3}$, 8.9$\times 10^{3}$) & 2.4$\times 10^{3}$ (2.3$\times 10^{3}$, 2.4$\times 10^{3}$) & 13115  (12379, 13782) &            ---            \\
PN 109         &              ---              & 6.1$\times 10^{3}$ (5.9$\times 10^{3}$, 6.2$\times 10^{3}$) & 12481  (11681, 13097) &            ---            \\

\hline
\multicolumn{5}{l}{
$a$ In parentheses we give the   15th and 85th percentiles  computed by the Monte Carlo simulations.} \\
\end{tabular}
\end{table*}


\begin{table*} [h!]
\caption{Ionic abundances for the PNe and compact \hii\ regions in   NGC\,300 (table  available electronically only).}  
\label{tab:ion}\centering\begin{tabular}{lrrrrrrrrrr}
\hline \hline 

 object   &   He$^{+}$    &  He$^{++}$       & O$^{+}_{3727}$  & O$^{+}_{7325}$   &   O$^{++}$     &    N$^{+}$    &   Ne$^{++}$     &     S$^{+}$    &   S$^{++}_{9069}$   &   Ar$^{++}$ \\
           &      &      &     ($ 10^6$) &     ($ 10^6$) &     ($10^6$) &      ($10^6$) &      ($10^6$)&      ($10^6$)  &      ($10^6$)&      ($10^6$) \\
           \hline

 H\,{\sc ii}   2    &   0.104  &   0.000     &        107.9 &   79.0 &   36.6 &     5.34   &    3.22   &    0.82   &   4.00  &  1.11 \\
 H\,{\sc ii}   5    &   0.271  &   0.000     &        122.0 &  115.8 &  109.9 &     7.31   &   18.38   &    0.92   &   4.91  &  1.05 \\
 H\,{\sc ii}   8    &   0.079  &   0.000     &        110.8 &   80.8 &   51.1 &     6.76   &    5.01   &    0.80   &   6.24  &  1.54 \\
 PN   12    &   0.000  &   0.053     &         14.7 &    0.0 &  271.5 &     2.04   &   34.47   &    0.00   &   0.00  &  0.00 \\
 PN   14    &   0.079  &   0.021     &         17.2 &    0.0 &   99.7 &     4.71   &   19.58   &    0.48   &   2.48  &  0.71 \\
 PN   22    &   0.106  &   0.000     &          0.0 &   15.3 &  108.3 &     2.02   &   22.33   &    0.20   &   1.16  &  0.67 \\
 PN   24    &   0.054  &   0.040     &         11.9 &    0.0 &  176.1 &     2.86   &   34.29   &    0.00   &   4.39  &  1.00 \\
 PN   25    &   0.111  &   0.025     &         61.7 &   25.6 &  195.9 &    11.66   &   43.67   &    0.94   &   2.52  &  1.20 \\
 PN   35    &   0.110  &   0.014     &         17.6 &   12.9 &  144.8 &     9.85   &   28.57   &    0.42   &   1.44  &  0.73 \\
 H\,{\sc ii}  39    &   0.054  &   0.000     &         49.8 &  150.4 &  124.2 &   113.05   &   17.93   &    1.64   &   2.37  &  0.55 \\
 PN   40    &   0.093  &   0.003     &          5.2 &   66.7 &  170.0 &     3.08   &   25.65   &    0.00   &   2.09  &  0.68 \\
 PN   45    &   0.119  &   0.000     &         31.1 &    0.0 &  300.8 &     4.34   &   61.45   &    0.51   &   1.58  &  0.65 \\
 PN   48    &   0.136  &   0.000     &          5.1 &   40.8 &  191.7 &   111.11   &   33.36   &    0.31   &   2.31  &  1.32 \\
 PN   51    &   0.118  &   0.000     &         15.4 &   33.5 &  157.1 &     4.30   &   26.24   &    0.00   &   2.40  &  1.07 \\
 PN   54    &   0.099  &   0.004     &         74.9 &   91.5 &  202.8 &   335.93   &   56.10   &    1.03   &   3.96  &  1.83 \\
 H\,{\sc ii}  57    &   0.093  &   0.002     &        138.3 &  177.5 &  206.5 &   118.14   &   67.20   &    3.04   &   4.73  &  1.80 \\
 PN   58    &   0.085  &   0.030     &         29.9 &   43.3 &  186.0 &   223.43   &   42.75   &    0.73   &   2.80  &  1.39 \\
 PN   63    &   0.093  &   0.000     &         69.3 &   80.1 &  121.8 &     5.42   &   14.40   &    0.53   &   6.96  &  1.46 \\
 PN   65    &   0.088  &   0.014     &          3.0 &    0.0 &  194.9 &     1.20   &   34.33   &    0.00   &   0.00  &  0.00 \\
 PN   66    &   0.109  &   0.013     &          9.2 &    0.0 &  196.6 &     2.66   &   33.35   &    0.16   &   1.05  &  0.56 \\
 PN   69    &   0.107  &   0.000     &          4.8 &   56.1 &  149.8 &     2.64   &   29.58   &    0.16   &   1.63  &  0.64 \\
 PN   74    &   0.113  &   0.000     &         43.6 &  129.5 &  169.4 &     4.56   &   27.52   &    0.43   &   6.06  &  1.88 \\
 H\,{\sc ii}  87    &   0.092  &   0.000     &         82.3 &   66.3 &   50.0 &     5.90   &    6.41   &    0.62   &   4.18  &  0.94 \\
 PN   88    &   0.000  &   0.043     &         12.6 &    0.0 &  110.6 &     6.30   &   20.86   &    1.01   &   0.00  &  0.94 \\
 H\,{\sc ii}  90a   &   0.101  &   0.000     &         93.7 &   91.2 &   77.7 &     4.96   &   10.98   &    0.62   &   5.42  &  1.33 \\
 H\,{\sc ii}  90b   &   0.101  &   0.000     &        119.4 &  105.1 &   78.6 &     3.99   &   10.85   &    0.45   &   3.35  &  1.33 \\
 PN   91    &   0.081  &   0.041     &          7.6 &    0.0 &  137.2 &     2.13   &   18.05   &    0.09   &   1.23  &  0.51 \\
 PN   96    &   0.044  &   0.017     &          3.1 &    0.0 &  124.1 &     1.31   &   23.96   &    0.00   &   0.00  &  0.00 \\
 PN   97    &   0.106  &   0.000     &          0.0 &    0.0 &  194.1 &     0.00   &   34.12   &    0.00   &   0.00  &  0.87 \\
 H\,{\sc ii}  102    &   0.085  &   0.000     &         91.7 &  101.4 &   47.8 &     4.01   &    5.55   &    0.80   &   2.80  &  0.97 \\
 PN  103    &   0.115  &   0.012     &         14.4 &   19.0 &  283.2 &     4.28   &   37.30   &    0.18   &   1.24  &  0.85 \\
 PN  104    &   0.097  &   0.048     &         13.0 &    0.0 &   96.3 &     6.62   &   22.12   &    0.32   &   0.94  &  0.58 \\
 PN  108    &   0.093  &   0.010     &          0.0 &   12.1 &  176.9 &     3.48   &   28.46   &    0.37   &   1.44  &  0.72 \\
 PN  109    &   0.000  &   0.017     &          0.0 &    0.0 &  180.7 &     0.00   &   40.91   &    0.00   &   0.00  &  0.00 \\\hline
\end{tabular}
\end{table*}

The ionic abundances were derived from the following lines (when observed): \Hei, \Heii, 
\Oii, \Oiii, \Nii, \Neiii, \Sii, \Siii, and \Ariii\
with respect to H$\beta$.  In the few cases where the \Oii\ line was not observed but the \Oiitoneb\
was, we used the latter to derive the \Op\ abundance. When only an upper limit was available for a line intensity, the intensity was put to zero for the calculations. We checked that the errors on the abundances resulting form this procedure are smaller than the uncertainties presented  in Sect. \ref{sec:uncertainties}. 

For PNe, we assumed that the temperature of all the ions is equal to $T$\oiii, the temperature given by \rOiii\footnote{For one of our objects, PN 20,  no direct measurement of the electron temperature is possible, so no abundances have been calculated.}. This is, of course, not entirely correct, as temperature gradients are expected to occur inside the nebulae. One could be tempted to use the empirical relation found to exist between $T$\oiii/$T$\nii\ and \ion{He}{ii} 4686/\Hb\ by Kingsburgh \& Barlow (1994) or Wang \& Liu (2007). However, this relation shows an important spread and its trend is opposite to that seen in the sample of Pe\~na et al. (2001) and Garc{\'{\i}}a-Rojas et al. (2012). In the few cases where the temperature from \rNii\ was available, we decided not to use it for abundance determination, the error bars being too large to present any advantage over the  temperature from \rOiii. For the compact \hii\ regions in our sample, we adopted the same procedure to relate the temperatures of the different ions as described in Bresolin et al. (2009), since there are many theoretical and observational arguments to justify it, as can be seen in Fig. 2 of Bresolin et al (2009). \footnote{This way of proceeding assumes that we already know which of our objects are PNe and which are compact \hii\ regions, and we had to iterate after making these decisions as explained in Sect. \ref{sec:distinguish}.}

In many cases, the only information available on the electron density comes from the \rSii\ line ratio. We adopt this value as representative for the entire nebula. A good evaluation of the density is crucial for \Op\ and \Sp\ but not so much for the other ions, since the critical density for collisional deexcitation of \Oiii, \Nii, \Neiii, \Siii, and \Ariii\ is rather high. There are a few cases where the density could be obtained only from  \rAriv, and this was adopted for all the ions. In the cases were the densities could be obtained from both \rSii\ and \rAriv\, we chose the density from \rSii\ to represent the nebula, since, as just mentioned, it is the density in the low ionization zone which is more critical for abundance determinations. However, there are some objects for which the error bar on the density from \rAriv\ is significantly smaller than that from \rSii\ or the density derived from \rAriv\ is more reasonable. For such cases (i.e. for objects 25, 35, 91, 103) we used the density obtained from \rAriv. When the spectra did not give any information at all on density (6 cases), we adopted a value of 3000\cmcub\ for PNe, and 100\cmcub\ for \hii\ regions. We note that, for objects 91 and 103, the only ones where the density could be derived from both \rSii\ and \rAriv\ with a certain degree of reliability, the density from \ariv\ is smaller than that of \sii\ by a factor 5 and 3, respectively. It is not clear whether this reflects a true difference in density between the high and low excitation zones. Given that, in the one hundred Galactic planetary nebulae observed by Wang et al. (2004), the differences between the two diagnostics are not that large, we are rather inclined to consider that our error bars might have been underevaluated. Note that in both cases, the \sii\  lines are extremely weak.

Table \ref{tab:ion} lists the ionic abundances determined for all our objects.

\subsection{Elemental abundances}
\label{sec:abund}


\begin{table*} [h!]
\caption{Computed abundances for the PNe and \hii\ regions in NGC\,300$^a$ (table  available electronically only).}
\label{tab:abundances}\centering\begin{tabular}{lcccccc}
\hline \hline 
\smallskip
 object     &         He/H           &     12+log O/H      &    12+log N/H       &    12+log Ne/H      &     12+log S/H      &    12+log Ar/H      \\
\hline
H\,{\sc ii}  2         & 0.104 (0.090,0.116) & 8.16 (8.02,8.38) & 6.92 (6.73,7.07) & 6.98 (6.42,7.24) & 6.66 (6.60,6.84) & 6.12 (6.00,6.35) \\
H\,{\sc ii}  5         & 0.271 (0.261,0.279) & 8.37 (8.27,8.48) & 7.22 (7.16,7.30) & 7.52 (7.40,7.68) & 6.58 (6.52,6.65) & 6.06 (5.99,6.15) \\
H\,{\sc ii}  8         & 0.079 (0.070,0.086) & 8.21 (8.05,8.47) & 7.07 (6.97,7.23) & 7.09 (6.91,7.40) & 6.85 (6.77,7.01) & 6.26 (6.14,6.45) \\
PN  12         & 0.053 (0.045,0.063) & 8.46 (8.30,8.64) & 7.60 (7.21,7.85) & 7.56 (7.36,7.81) &        ---          &
    ---        \\
PN  14         & 0.100 (0.095,0.112) & 8.14 (8.07,8.46) & 7.57 (7.43,7.77) & 7.43 (7.35,7.73) & 6.59 (6.53,6.80) & 6.12 (6.09,6.21) \\
PN  20         & 0.000 (0.000,0.000) &        ---          &   ---     &        ---          &     ---                  &         ---         \\

PN  22         & 0.106 (0.082,0.126) & 8.09 (8.00,8.20) & 7.21 (7.04,7.50) & 7.41 (7.29,7.54) & 6.10 (6.05,6.23) & 6.10 (6.03,6.18) \\
PN  24         & 0.094 (0.075,0.110) & 8.45 (8.37,8.56) & 7.83 (7.77,7.95) & 7.73 (7.64,7.86) & 6.71 (6.66,6.78) & 6.28 (6.18,6.35) \\
PN  25         & 0.135 (0.121,0.148) & 8.50 (8.44,8.58) & 7.77 (7.72,7.85) & 7.81 (7.73,7.92) & 6.58 (6.51,6.67) & 6.35 (6.30,6.42) \\
PN  35         & 0.124 (0.112,0.132) & 8.27 (8.21,8.33) & 8.01 (7.95,8.07) & 7.54 (7.47,7.62) & 6.39 (6.29,6.47) & 6.13 (6.07,6.20) \\
H\,{\sc ii} 39         & 0.054 (0.046,0.062) & 8.24 (8.09,8.40) & 7.69 (7.57,7.79) & 7.36 (7.17,7.56) & 6.64 (6.57,6.73) & 5.76 (5.65,5.88) \\
PN  40         & 0.096 (0.064,0.128) & 8.25 (8.15,8.37) & 8.02 (7.92,8.19) & 7.43 (7.31,7.58) & 6.56 (6.45,6.68) & 6.10 (6.03,6.18) \\
PN  45         & 0.119 (0.100,0.140) & 8.52 (8.34,8.83) & 7.66 (7.47,7.91) & 7.83 (7.61,8.18) & 6.83 (6.57,7.19) & 6.09 (5.73,6.34) \\
PN  48         & 0.136 (0.108,0.162) & 8.29 (8.20,8.43) & 8.63 (8.40,8.81) & 7.53 (7.43,7.70) & 6.75 (6.64,6.89) & 6.39 (6.32,6.50) \\
PN  51         & 0.118 (0.106,0.133) & 8.24 (8.16,8.31) & 7.68 (7.63,7.73) & 7.46 (7.37,7.55) & 6.52 (6.45,6.58) & 6.30 (6.23,6.37) \\
PN  54         & 0.103 (0.093,0.111) & 8.45 (8.40,8.52) & 8.14 (8.08,8.17) & 7.90 (7.83,7.97) & 6.70 (6.65,6.76) & 6.53 (6.46,6.60) \\
H\,{\sc ii} 57         & 0.095 (0.085,0.103) & 8.54 (8.43,8.65) & 7.72 (7.64,7.80) & 8.00 (7.86,8.15) & 6.90 (6.84,6.97) & 6.29 (6.21,6.36) \\
PN  58        & 0.116 (0.106,0.128) & 8.42 (8.38,8.48) & 8.32 (8.28,8.35) & 7.78 (7.73,7.85) & 6.72 (6.68,6.77) & 6.41 (6.37,6.46) \\
PN  63         & 0.093 (0.080,0.102) & 8.28 (8.06,8.51) & 7.17 (7.03,7.30) & 7.35 (6.99,7.66) & 6.92 (6.80,7.05) & 6.44 (6.30,6.58) \\
PN  65         & 0.102 (0.092,0.114) & 8.34 (8.28,8.40) & 7.94 (7.51,8.23) & 7.58 (7.52,7.66) &        ---          &
         ---         \\
PN  66         & 0.121 (0.109,0.129) & 8.35 (8.31,8.39) & 7.80 (7.71,7.85) & 7.57 (7.54,7.63) & 6.45 (6.36,6.52) & 6.02 (5.94,6.07) \\
PN  69         & 0.107 (0.095,0.119) & 8.19 (8.13,8.29) & 7.92 (7.78,8.04) & 7.48 (7.40,7.60) & 6.50 (6.45,6.58) & 6.08 (6.03,6.15) \\
PN  74         & 0.113 (0.101,0.125) & 8.33 (8.16,8.63) & 7.35 (7.25,7.52) & 7.54 (7.33,7.89) & 6.93 (6.84,7.08) & 6.55 (6.44,6.76) \\
H\,{\sc ii} 87         & 0.092 (0.083,0.100) & 8.12 (8.04,8.21) & 7.06 (6.95,7.12) & 7.14 (6.87,7.27) & 6.73 (6.71,6.79) & 6.04 (5.94,6.10) \\
PN  88         & 0.043 (0.038,0.047) & 8.09 (8.00,8.20) & 7.79 (7.70,7.88) & 7.37 (7.25,7.50) & 7.10 (7.00,7.19) & 6.24 (6.17,6.33) \\
H\,{\sc ii} 90a        & 0.101 (0.094,0.108) & 8.23 (8.15,8.37) & 7.04 (6.97,7.13) & 7.31 (7.20,7.49) & 6.77 (6.71,6.86) & 6.17 (6.09,6.29) \\
H\,{\sc ii} 90b        & 0.101 (0.018,0.208) & 8.30 (8.18,8.44) & 6.90 (6.83,7.00) & 7.35 (7.20,7.51) & 6.50 (6.44,6.58) & 6.18 (6.11,6.27) \\
PN  91         & 0.122 (0.110,0.137) & 8.26 (8.22,8.30) & 7.71 (7.66,7.77) & 7.40 (7.34,7.46) & 6.57 (6.45,6.67) & 5.98 (5.90,6.04) \\
PN  96         & 0.062 (0.048,0.074) & 8.20 (8.15,8.27) & 7.81 (7.63,7.98) & 7.49 (7.41,7.56) &        ---          &
        ---          \\
PN  97         & 0.106 (0.089,0.117) & 8.29 (8.13,8.45) &        ---          &
 7.53 (7.34,7.72) &        ---          &
 6.21 (5.91,6.40) \\
H\,{\sc ii} 102         & 0.085 (0.076,0.093) & 8.14 (8.03,8.27) & 6.87 (6.70,6.96) & 7.11 (6.73,7.28) & 6.51 (6.45,6.59) & 6.05 (5.90,6.17) \\
PN 103         & 0.128 (0.114,0.144) & 8.34 (8.29,8.39) & 7.81 (7.76,7.87) & 7.63 (7.58,7.69) & 6.43 (6.36,6.51) & 6.20 (6.16,6.25) \\
PN 104         & 0.145 (0.131,0.160) & 8.16 (8.09,8.22) & 7.86 (7.80,7.92) & 7.52 (7.44,7.58) & 6.25 (6.14,6.32) & 6.04 (5.93,6.14) \\
PN 108         & 0.103 (0.092,0.112) & 8.31 (8.23,8.38) & 7.76 (7.57,7.94) & 7.51 (7.43,7.60) & 6.61 (6.43,6.75) & 6.13 (5.99,6.24) \\
PN 109         & 0.017 (0.015,0.020) & 8.26 (8.20,8.35) &        ---          &
 7.61 (7.54,7.71) &        ---          &   ---   \\
\hline
\multicolumn{7}{l}{$^a$ In parentheses are given the  15th and 85th percentiles  computed by the Monte Carlo simulations} \\
\end{tabular}
\end{table*}

To obtain the elemental abundances from ionic abundances, one must correct for unseen ions. The present situation with ionization correction factors (ICFs) is far from satisfactory. In view of the lack of better options, for PNe we use the ICFs from Kingsburgh \& Barlow (1994), which were obtained from detailed models for 10 PNe. However, these models were never published, and it is not clear whether the modelled objects are similar to the ones studied in the extragalactic case, where it is the brightest PNe that are chosen for spectroscopic studies. In addition, our spectra encompass the entire spatial extent of the PNe, while those of  Kingsburgh \& Barlow (1994) correspond to only small zones of the objects. For \hii\ regions, we follow the same policy as Bresolin et al. (2009). It uses the ICFs from Izotov et al. (2006), which are based on sequences of photoionization models aimed at  reproducing the observed properties of giant \hii\ regions. The only exception is represented by neon, for which we adopt  Ne/O$\,=\,$\Nepp/\Opp, since the ICF proposed by Izotov et al. (2006) leads to a trend in Ne/O as a function of excitation. 
Lastly, in the case of helium we do not attempt to use any ICF, since there is no reliable way to correct for neutral helium in our objects. 

Table \ref{tab:abundances} lists the nominal abundances of O, N, Ne, S, Ar with respect to H, as well as the sum of the abundances of \Hep\ and \Hepp\ with respect to H. We also indicate the 15 and 85 percentiles, obtained from our Monte Carlo analysis of errors, as explained in the next section.  

Since we wish to compare the abundances of PNe with those of \hii\ regions, we recomputed the abundances of the giant \hii\ regions from Bresolin et al. (2009) with the same atomic parameters and procedures as for the compact \hii\ regions presented in this paper.   Our abundances are in general very similar to the ones presented in Bresolin et al. (2009), but differences can amount to 0.1- 0.2 dex in the objects with the highest abundances, due to the different atomic data employed. 

\begin{figure}
\includegraphics[width=\columnwidth]{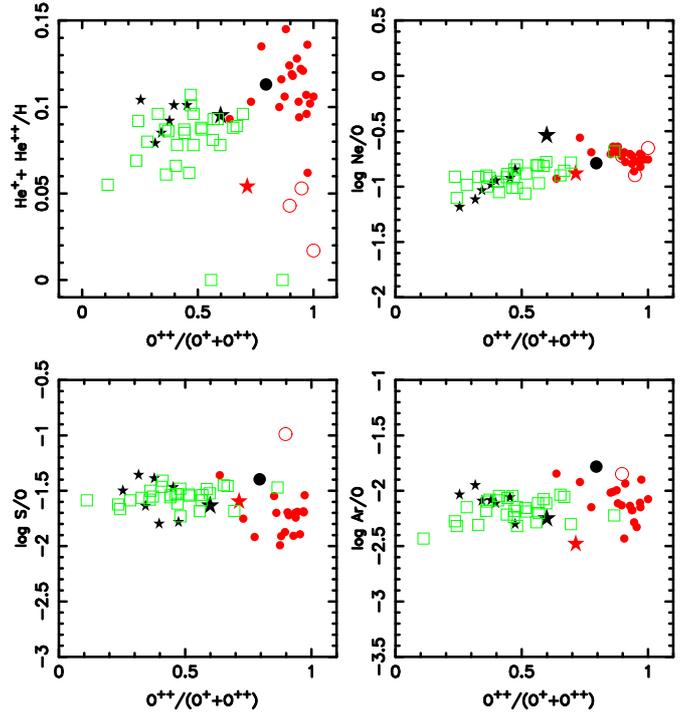} 
\caption{Various abundance ratios as a function of \Opp/(\Op\ + \Opp) for PNe and \hii\ regions in NGC 300. PNe are represented by circles, compact \hii\ regions by stars and giant \hii\ regions by squares. The big red star is object no. 39, the big black circle is object no. 74 (see Sect. \ref{sec:distinguish}), and the big black star is object no. 57 (see Sect. \ref{sec:abund}).  The objects represented by open circles are objects  12, 88 and 109 (see Sect. \ref{sec:uncertainties}).  \label{fig:icfs}}
\end{figure}

To test  the adopted ICFs, we plot in Fig. \ref{fig:icfs} the values of Ne/O, S/O and Ar/O as a function of \Opp/(\Op+\Opp).  Throughout the paper, PNe are represented by circles, compact \hii\ regions by stars, and the giant \hii\ regions from Bresolin et al. (2009) by green squares. No significant trend with excitation is seen for Ar/O and  S/O, although for the latter we find a tendency for PNe to give a lower value than \hii\ regions. We are here witnessing the sulfur problem discussed at lenght by Henry et al. (2012). For Ne/O, there is no trend with excitation in the case of PNe, but a slight one for \hii\ regions -- both giant and compact ones -- (which had already been noted by Bresolin et al. 2009 for giant \hii\ regions and probably implies that the ICFs for Ne in \hii\ regions should be revised). 

In the top left panel of Fig. \ref{fig:icfs} we plot the values of (\Hep+\Hepp)/H as a function of \Opp/(\Op+\Opp). Some of these values are very low, clearly  indicating the presence of neutral helium. There is no reliable way to correct for that, since the ionization of helium is mainly governed by the temperature of the ionizing star, while that of the heavy elements also depends on the ionization parameter. 

\subsection{Uncertainties}
\label{sec:uncertainties}

\begin{figure}
\includegraphics[width=5 cm,trim=0mm 40mm 0mm 0mm, clip=true]{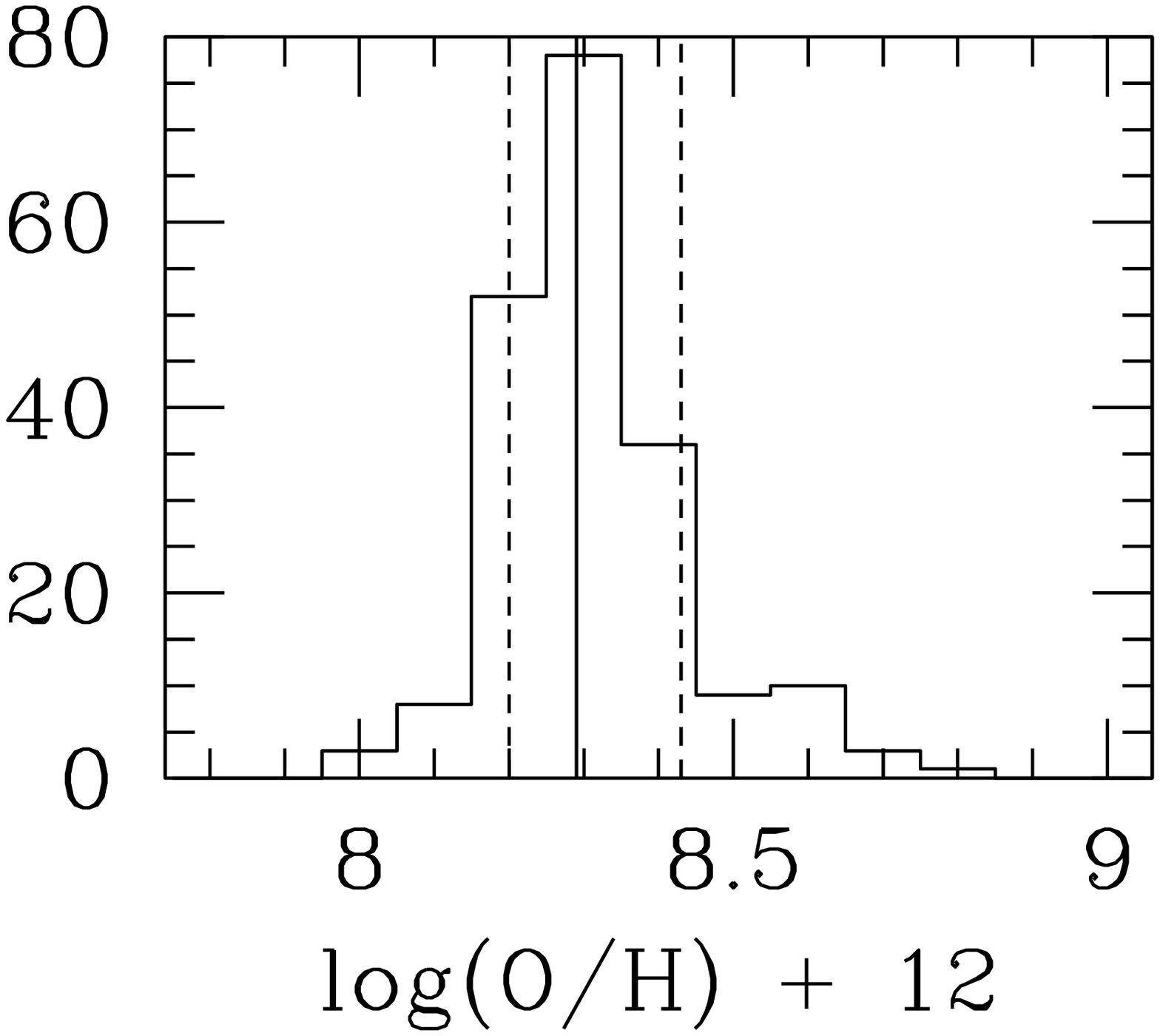} 
\caption{Histogram of log O/H +12 values for object PN 48 from the Monte Carlo simulation. The nominal value is indicated by the continuous line, the 15 and 85 percentiles by the dashed lines. \label{fig:histogram}}
\end{figure}

To estimate the uncertainties in the derived abundances we start from the uncertainties in the line intensities listed in Table  \ref{tab:intensities}.  Those were estimated as explained in section 2.2, and include the uncertainties involved in the dereddening process.  We use a Monte Carlo procedure to follow the error propagation in the determination of the plasma parameters and the abundances. We assume for each line flux a gaussian distribution centered on the flux effectively measured and having a dispersion equal to the estimated flux uncertainty, and given in Table \ref{tab:intensities}\footnote{This is perhaps not the most realistic distribution, but it is convenient. Wesson et al. (2012) have used more sophisticated distributions, but the strong biases they claim for line intensities with signal-to-noise ratios $\le4$ are not supported by observations of line ratios such as  [O\,{\sc iii}]\,$\lambda$4959/5007 or [N\,{\sc ii}]\,$\lambda$6548/6484 which, on average, are consistent with the values predicted by atomic physics (see, e.g,. Fig.~10 of Bresolin et al.~2005) }. For each line, we considered 200 independent realizations, and estimated the error bars  by taking the 15 and 85 percentiles of the final results\footnote{Whenever this procedure lead to a negative value for a line intensity, which may happen principally for low signal-to-noise cases, it was replaced by one tenth of the measured value. Since we use percentiles and not standard deviations to estimate the uncertainties, such a way of dealing with the problem is acceptable.}. The reason for considering percentiles and not one sigma deviations is that the final distributions on abundances can be highly non-gaussian, as shown in Fig. \ref{fig:histogram} for the case of PN 48, and we are in fact interested in the range of values in which there is a 70\% chance for the real value to be found (nb: in the case of a gaussian distribution, an interval of $\pm 1 \sigma$ corresponds to a probability of 68.3\%). Note that the nominal values of the abundances listed in Table \ref{tab:abundances} do not necessarily correspond to the median or the most probable values of the Monte Carlo simulation. Our estimation for the final uncertainties on the abundances does not include uncertainties on ICFs, for which we have no handle. This is unfortunately the situation for all abundance determinations using ICFs that have been published so far.

For PN 97 and 109 no lines from \Op\ are measured, therefore the abundance of that ion cannot be estimated. The total oxygen abundance should however not be drastically affected since in that case the contribution of \Op\ to the O/H abundance ratio is expected to be small.  In PN 12, 88 and 109  the \Hei\ line is not observed and O/H is obtained simply by adding \Opp\ and \Op\ (when measured), so that the computed O/H is a lower limit to the true oxygen abundance, since it does not account for the presence of higher ionization species of oxygen. In all the plots in this paper these three objects are represented by open symbols, and are not considered in the computation of the radial abundance gradients in Sect. \ref{sec:gradients}. On the other hand, PN 97 is represented with the same symbols as the rest of PNe and is included in the computation of the gradients in Sect. \ref{sec:gradients}.

Concerning  \hii\ regions, Stasi\'nska (2005) made the point that abundances derived using temperature-based methods in metal-rich nebulae are prone to strong errors, due to important temperature gradients with the nebulae. From the behavior of the radial variation of O/H in the giant \hii\ regions observed by Bresolin et al. (2009), it does not seem that these objects belong to the abundance regime where such effects appear (except perhaps the two innermost \hii\ regions):  the radial variation of O/H is rather smooth, and the dispersion can be explained  by observational errors only.


\section {How to distinguish PNe from \hii\ regions in NGC\,300}
\label{sec:distinguish}

\begin{figure} [h!]
\includegraphics[width=\columnwidth]{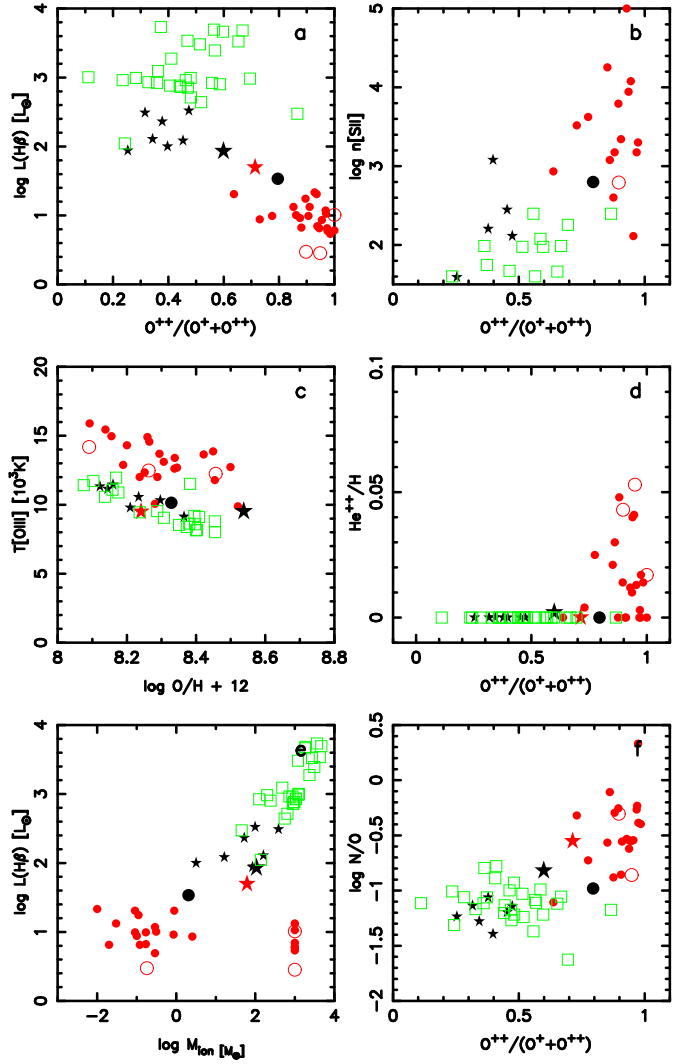} 
\caption{Diagrams illustrating observed properties that allow to differentiate between PNe and \hii\ regions in NGC 300 (see Sect. \ref{sec:distinguish}). The symbols have the same meaning as in Fig.  \ref{fig:icfs}. The few PNe for which the electron density cannot be measured have been placed at log $M_{\rm ion} = 3$. \label{fig:PNeHII}}
\end{figure}

In Pe\~na et al. (2012), the candidate PNe were defined as point-like objects seen in \oiii\ emission but not in the continuum. If the objects were seen in the continuum, they were considered as candidate  \hii\ regions. The spectroscopic data presented in this paper allow us to revise our classification on a more solid basis. 

In Fig. \ref{fig:PNeHII} we show a  few diagrams that are useful for this purpose.   In this figure, as well as in all the similar figures in this paper, object number 74  is represented by a big black circle, and object number 39 is represented by a big red star. From our photometric analysis, initially object 74 was considered as a compact \hii\ region and object 39 as a PN but, as we will see, spectroscopic data argue for the opposite. In all the diagrams of Fig. \ref{fig:PNeHII}, we can see a clear segregation between candidate \hii\ regions and candidates PNe.  The \Hb\ luminosities of the latter tend to be lower that those of candidate \hii\ regions. As expected, many candidate PNe show \Heii\ emission while none of the  \hii\ regions do. Many candidate PNe  have higher N/O ratios than \hii\ regions, similarly to what is found for well-known PNe in the Milky Way (Henry et al. 2000). PNe tend to have a higher  \oiii\ temperature at a given O/H since, on average, the effective temperatures of their exciting stars are much higher that those of massive stars that ionize \hii\ regions. For the same reason, PNe tend to have a higher excitation, as measured by the \Opp/(\Op + \Opp) ratio. They also tend to have higher densities. Indeed, PNe with densities similar to those of \hii\ regions would be too faint to be detected at the distance of NGC\,300, since the \Hb\ luminosity is proportional to the gas density times the nebular mass which, in the case of PNe is at most of the order of a few  solar masses. For a temperature of $10^4$K, relevant for our sample, the  mass of ionized gas is given by  
\begin{equation}
\label{eq:mass}
M_{\rm ion} = 37.5   L(\Hb) / n
\end{equation}
where $L(\Hb)$ and $M_{\rm ion}$ are in solar units, and $n$ is the mass-weighted average electron density in cm$^{-3}$. In practise, we use  the \sii\ or \ariv\ ratios to estimated this density.

The $L(\Hb)$ versus $M_{\rm ion}$ relation (Fig. \ref{fig:PNeHII} bottom left) clearly shows the sequence: PNe -- compact \hii\ regions -- giant \hii\ regions. Of course, the estimates of $M_{\rm ion}$ are coarse, the densities being uncertain,  as seen in Table \ref{tab:plasma}, but even considering those uncertainties, the trend is clearly present. 
In all the diagrams of Fig. \ref{fig:PNeHII} objects number 39 and number 74 appear at the junction between PNe and compact \hii\ regions. So why do we consider object 74 as a PN and object number 39 as an \hii\ region? There are strong arguments for that. If object 74 were a compact \hii\ region, from its \Hb\ luminosity one would deduce that the effective temperature of the ionizing star would be about 30\,000\,K, using the relation between the Lyman continuum output and the effective temperature of main sequence stars from Martins et al. (2005). But the observed value of \Opp/(\Op + \Opp)  for this object is 0.79, clearly incompatible with such a low effective temperature. On the other hand, for  object 39, using the equation above, we find an ionized mass of around 100\,\msun, clearly incompatible with a planetary nebula. 

Such sanity checks on the classification of point-like extragalactic nebulae  are very important. In the present case, only 2 objects out of 35 would have been misclassified without the use of spectroscopic data. But the proportion could be much larger in other studies, especially if the preselection is done with a telescope too small to detect the continuum emission of an O-type star. 


\section {Abundance patterns of PNe and \hii\ regions in NGC\,300}
\label{sec:patterns}

\begin{figure}
\includegraphics[width=\columnwidth]{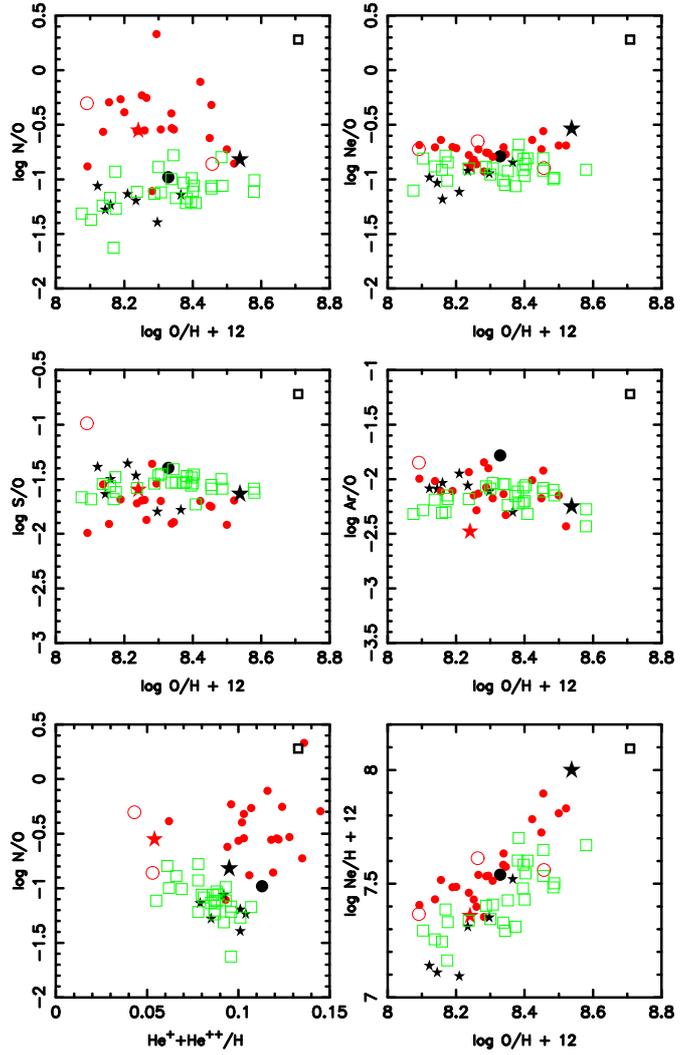} 
\caption{Abundance patterns of PNe and \hii\ regions in NGC 300. The symbols have the same meaning as in Fig.  \ref{fig:icfs}.  \label{fig:abpatterns}}
\end{figure}

If Fig. \ref{fig:abpatterns} we present some classical diagrams comparing the abundance patterns of PNe and   \hii\  regions in NGC 300. 

The most evident feature is the systematically larger N/O ratio in PNe relative to \hii\ regions. Such a behavior is observed in other galaxies like, for example, the Magellanic Clouds (Leisy \& Dennefeld 2006), NGC 6822 (Hern\'andez-Mart\'inez et al. 2009)  or M 33 (Bresolin et al. 2010). As seen in panel e of Fig.  \ref{fig:abpatterns}, N/O appears to be correlated with the helium abundance as estimated from (\Hep\ + \Hepp)/H (discarding the PNe PN 12 and 88 for which no \hei\ line is measured). This is again similar to what was found in previous studies of other galaxies and is in agreement with the idea that the observed enhancement of nitrogen is due to the second dredge-up, which brings  CNO-cycle processed material to the stellar surface.

The Ne/O ratio is very constant among both PNe and \hii\ regions. There is a small offset (by about 0.2 dex) seen in Panels b and f between the two types of objects. Such an offset has already been shown to occur in other galaxies (e.g., Bresolin et al. 2010). It is difficult to say whether it  tells us  something about the physics of these objects rather than simply being due to inadequate ICFs (see however Sect. \ref{sec:discussion}).  Note that the compact \hii\ region H\,{\sc ii} 57 has surprisingly high value of Ne/O (0.29) for a reason that we do not understand. 

The Ar/O values also do not show any trend with O/H, both for \hii\ regions and PNe. For the latter objects, the dispersion is larger than for Ne/O, probably both because the \Ariii\ lines are weaker than  the \Neiii\ lines and because the ICF used for argon is extremely coarse.

For sulfur, we find S/O ratios systematically lower for PNe than for \hii\ regions, similar to what was found by other authors in many samples of PNe (see Henry et al. 2012 for a recent discussion).


\section {Radial abundance gradients in NGC\,300}
\label{sec:gradients}

\begin{figure*}  
\includegraphics[width=\textwidth]{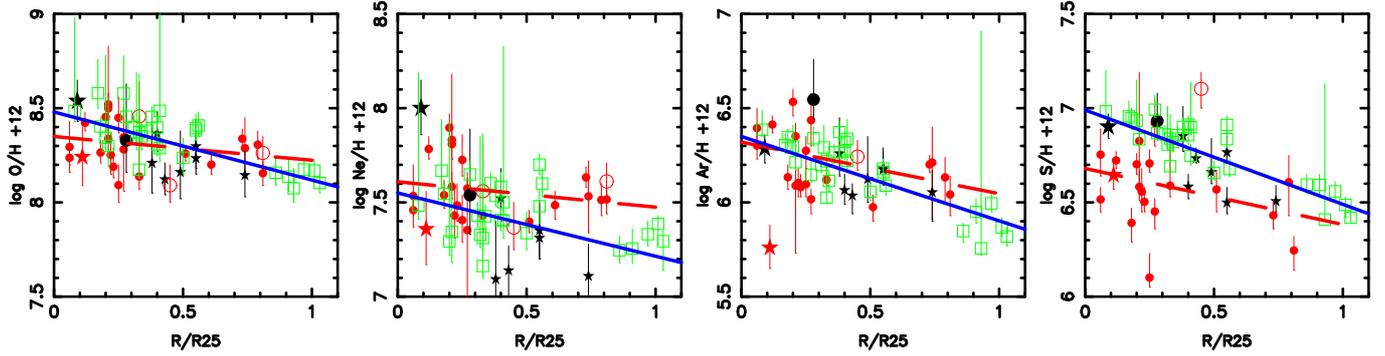} 
\caption{Abundances with respect to hydrogen as a function of the normalized galactocentric radius  $R/R_{25}$. The symbols have the same meaning as in Fig.  \ref{fig:icfs}. The continuous blue lines and the dashed red lines represent the straight line fits to the observations of \hii\ regions and of PNe, respectively. \label{fig:gradsH}}
\end{figure*}

\begin{figure*}  
\includegraphics[width=\textwidth]{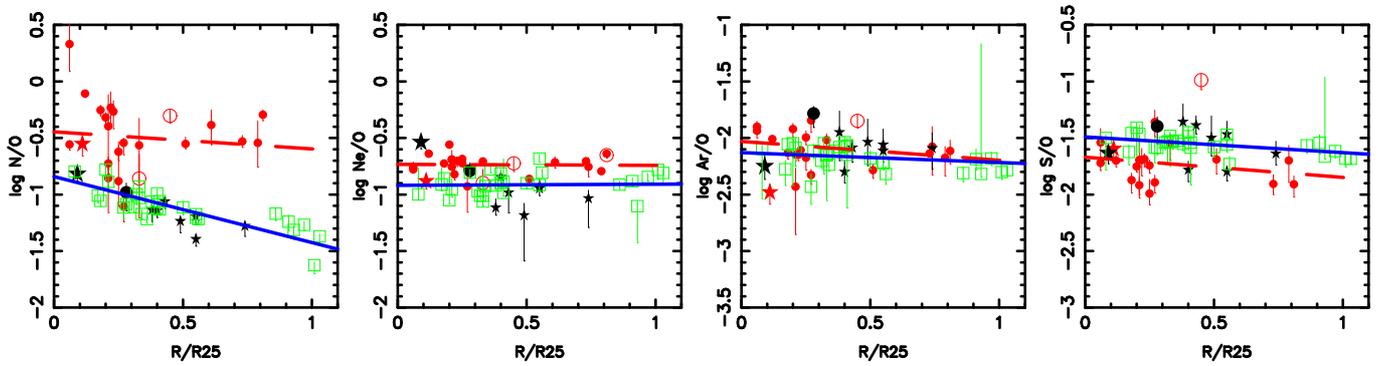} 
\caption{Abundances with respect to oxygen as a function of the normalized galactocentric radius  $R/R_{25}$. The layout of the figure is the same  as in Fig.  \ref{fig:icfs}.\label{fig:gradsO}}
\end{figure*}

\begin{table*} [h!]
\caption{Straight-line fitting of the radial variations of abundance ratios for the PNe and  \hii\ regions in   NGC\,300}   
\label{tab:resugrad}
\centering\begin{tabular}{ccccccc}
\hline \hline 

 $y$    &         &       $B$     &     $A$      & nb \\

\hline

12 + log O/H  & PNe  	 & $-0.126$ $\pm$ $0.0752$ & $8.35$ $\pm$ $0.0407$ & 22 \\
\medskip
12 + log O/H  & H\,{\sc ii}	 & $-0.361$ $\pm$ $0.0493$ & $8.48$ $\pm$ $0.0336$ & 37 \\

12 + log Ne/H & PNe  	 & $-0.135$ $\pm$ $0.0915$ & $7.61$ $\pm$ $0.0562$ & 22 \\
\medskip
12 + log Ne/H &H\,{\sc ii} 	 & $-0.335$ $\pm$ $0.116$ & $7.55$ $\pm$ $0.0680$ & 36 \\

12 + log Ar/H & PNe  	 & $-0.274$ $\pm$ $0.0916$ & $6.32$ $\pm$ $0.0546$ & 20 \\
\medskip
12 + log Ar/H & H\,{\sc ii}	 & $-0.448$ $\pm$ $0.0918$ & $6.35$ $\pm$ $0.0564$ & 37 \\

12 + log S/H  & PNe  	 & $-0.297$ $\pm$ $0.154$ & $6.68$ $\pm$ $0.0708$ & 19 \\
\medskip
12 + log S/H  & H\,{\sc ii} 	 & $-0.500$ $\pm$ $0.0569$ & $6.99$ $\pm$ $0.0341$ & 37 \\
\hline

log N/O  & PNe  	 & $-0.150$ $\pm$ $0.269$ & $-0.446$ $\pm$ $0.139$ & 21 \\
\medskip
log N/O  & H\,{\sc ii}	 & $-0.583$ $\pm$ $0.0908$ & $-0.842$ $\pm$ $0.0473$ & 37 \\

log Ne/O & PNe  	 & $-0.0095$ $\pm$ $0.0606$ & $-0.733$ $\pm$ $0.0278$ & 22 \\
\medskip
log Ne/O & H\,{\sc ii} 	 & $ 0.0099$ $\pm$ $0.0852$ & $-0.919$ $\pm$ $0.0481$ & 36 \\

log Ar/O & PNe  	 & $-0.167$ $\pm$ $0.0834$ & $-2.03$ $\pm$ $0.0553$ & 20 \\
\medskip
log Ar/O & H\,{\sc ii}	 & $-0.087$ $\pm$ $0.0722$ & $-2.13$ $\pm$ $0.0451$ & 37 \\

log S/O  & PNe  	 & $-0.178$ $\pm$ $0.110$ & $-1.67$ $\pm$ $0.0555$ & 19 \\
log S/O  & H\,{\sc ii} 	 & $-0.139$ $\pm$ $0.0416$ & $-1.49$ $\pm$ $0.0264$ & 37 \\
\hline
\end{tabular}
\end{table*}

We now focus on the radial behavior of the properties of PNe and \hii\ regions (considering compact \hii\ regions and giant \hii\ regions together). 

Figure \ref{fig:gradsH} shows the radial distribution of abundance ratios with respect to hydrogen, as a function of $R/R_{25}$, where $R_{25}$ is the radius out to the blue surface brightness level of  25 mag arcsec$^{-1}$, corrected for inclination and galactic extinction. This normalization of the radius is the most widely used since Vila-Costas \& Edmunds (1992), and allows one to soundly compare the radial behavior of abundance ratios in different galaxies. The symbols have the same meanings as in all the figures starting from Fig.  \ref{fig:icfs}. The estimated error in the line ratios is indicated for each object. One can see that, in the central zones, error bars for giant \hii\ regions are larger than those for PNe, despite the fact that the former are more luminous by about two orders of magnitude. This is due to the fact that metal-rich \hii\ regions have low electron temperatures, and an error in their estimated values strongly affects the resulting abundance. Note that the error bars reported in Fig. \ref{fig:gradsH} take into account only error propagation from the observed line fluxes, but consider that the scheme adopted to evaluate the temperatures of the different zones is perfectly valid, so the real uncertainties in the abundances with respect to hydrogen are likely higher, especially in \hii\ regions.

Figure \ref{fig:gradsO} is analogous to Fig. \ref{fig:gradsH} but shows abundance ratios with respect to oxygen. 

The radial trend of the abundances in compact \hii\ regions is similar to that in the giant \hii\ regions, although one may note a tendency for compact \hii\ regions to have systematicall lower oxygen and neon abundances than giant \hii\ regions at similar galactocentric radii. We do not see any convincing explanation for this fact.  Larger statistics would be needed to first ascertain whether the effect is real or not. On the other hand, PNe and \hii\ regions appear to have convincingly different radial behaviors. This is especially clear for abundance ratios involving O, N and Ne (and also S, but for this element, as noted  in Sect. \ref{sec:patterns} the abundances in PNe are uncertain).

 Table \ref{tab:resugrad} presents the results of straight-line fitting of various abundance ratios as a function of fractional galactocentric distance, $R/R_{25}$.  The fitting was done taking into account the uncertainties in abundance ratios, but discarding errors in radial distances.  For each abundance ratio, we performed one fitting for PNe, and one for \hii\ regions (merging compact and giant \hii\ regions). 
For each fitting, the table gives the values of $A$ (the intercept) and $B$ (the slope) in the equation
\begin{equation}
\label{ eq:gradient}
y = A + B \times R/R_{25},
\end{equation}
as well as the number of objects considered.

 Table \ref{tab:resugrad} clearly shows that the radial abundance gradients of PNe are significantly shallower than for \hii\ regions in the case of O/H, Ne/H, Ar/H and S/H. As a matter of fact, the abundance gradients for PNe are almost flat. We should note, however, that the abundance dispersion at any galactocentric radius is larger for PNe than for \hii\ regions. The abundance gradient that we find for O/H in our sample of \hii\ regions is slightly flatter  than found by Bresolin et al. (2009): $-0.36\pm 0.05$ as compared to $-0.41\pm 0.03$. As can be deduced from Fig. \ref{fig:gradsH}, this is due to the inclusion of the compact \hii\ regions. 
 
 We also note that the values of $A$ are smaller for PNe than for \hii\ regions, in a very significant way in the case of O/H.
 
Concerning the radial behavior of the abundance ratios with respect to oxygen we observe that Ne/O shows no gradient at all,  with the Ne/O value being systematically higher by almost 0.2 dex for PNe than for \hii\ regions at any galactocentric distance. There is no gradient in Ar/O for  \hii\ regions and a mild one for PNe (whose significance is however not obvious, given that the ICFs in the determination of the argon abundance are very crude, as mentioned earlier). Concerning S/O, the regression lines for PNe and \hii\ regions run almost parallel, with the PNe values being systematically lower than the \hii\ region ones by 0.2 dex. From those three ratios, given the gentle trends and small differences between PNe and \hii\ regions, and considering the uncertainties in the abundance derivation, it would be premature to draw any conclusion of astrophysical value. 

For N/O, on the contrary, the difference in the observed trends is spectacular. While for \hii\ regions the gradient in N/O is similar to the one in O/H (as already found in Bresolin et al. 2009), there is almost no gradient in the N/O values determined in PNe - just a large dispersion. Besides, at any galactocentric distance, almost all the observed values of N/O in our PNe are strongly above the 
 \hii\ region trend!  This implies that in most of the PNe in our sample the N/O ratio does not reflect that of the progenitor's star at birth, but has been heavily modified by dredge-up of nucleosynthesis products.

\section {Discussion}
\label{sec:discussion}

\begin{table*} [h!]
\caption{Typical properties of PNe and their progenitors}   
\label{tab:pnprop}
\centering\begin{tabular}{lrrrrrrrl}
\hline \hline 
 $M_{\rm i}$  [M$_{\odot}$]  &   1      &     2 & 2.5 & 3 & 4 & 5 & 6  & \\

 \hline
in  Fig. \ref{fig:npevol}    &   black      &  red    &  green  & blue  &  cyan & magenta &      &  \\
 \hline
  $t_{\rm MS}$ [yr]    &   $8.5 \times 10^{9}$     &   $1.0 \times 10^{9}$  & $5.4 \times 10^{8}$  & $3.2 \times 10^{8}$  & $1.5 \times 10^{8}$  & $8.8 \times 10^{7}$ & $5.7 \times 10^{7}$   & Ekstr\"om et al. (2012) $^{a}$ \\

$M_{\rm f}$  [M$_{\odot}$]   &   0.55      &    0.6 & 0.63 & 0.68 & 0.8 & 0.88 & 0.95  & Weidemann (2000)\\
\hline

$\Delta$ log N/O      &  0.00 & 0.30 & 0.36 & 0.38 & 0.44 & 0.68 &    &    Marigo (2001) $^{b}$ $Z =$ 0.02\\
\medskip
$\Delta$ log O/H     &    0.01 & 0.06 & 0.07 & 0.07 & 0.02 & 0.01 &    &     Marigo (2001) $^{b}$ $Z =$ 0.02\\

$\Delta$ log N/O      &   0.14   &  0.30   & 0.33 &  0.37 & 0.41  & 0.60 & 0.96  &    Karakas (2010) $^{b}$ $Z =$ 0.02\\
\medskip
$\Delta$ log O/H    &    0.00     &   0.00   &  $-0.01$ &  $-0.02$ & $-0.02$  & $-0.04$ & $-0.08$  &    Karakas (2010) $^{b}$ $Z =$ 0.02\\

$\Delta$ log N/O      &					0.21 &	0.41 &	0.47 &	0.54 &	0.55 & &	0.68 &	 Lagarde et al. (2012)		$^{d}$	$Z =$ 0.014\\		\medskip																														
$\Delta$ log O/H    &					0.01 &	0.00 &	$-0.01$ &	$-0.02$ &	$-0.03$ & &	$-0.03$ & 	Lagarde et al. (2012)		$^{d}$	$Z =$ 0.014\\	
					
$\Delta$ log N/O      &					0.38 &	0.50 &	0.56 &	0.59 & 0.62 & &	0.74	&		Lagarde et al. (2012)		$^{e}$	$Z =$ 0.014\\	
			
$\Delta$ log O/H    &		            0.02 & 	0.00 & $-0.02$ &	$-0.02$ & 	$-0.03$ & &	0.00 &  	Lagarde et al. (2012)		$^{e}$	$Z =$ 0.014\\	

\hline
$\Delta$ log N/O      &   0.03 & 0.19 & 0.26 & 0.28 & 1.05 & 1.20 &    &     Marigo (2001) $^{b}$ $Z =$ 0.008\\
\medskip
$\Delta$ log O/H     &    0.01 & 0.21 & 0.22 & 0.20 & 0.07 & 0.02 &    &     Marigo (2001) $^{b}$ $Z =$ 0.008\\

$\Delta$ log N/O      &   0.16   &  0.33   & 0.38 &  0.41 & 0.41  & 1.60 & 1.57  &    Karakas (2010) $^{\rm b}$ $Z =$ 0.008\\
$\Delta$ log O/H    &    0.00     &   0.00   &  $-0.01$ &  $-0.03$ & $-0.01$  & $-0.12$ & $-0.22$  &  Karakas (2010) $^{b}$ $Z =$ 0.008\\
\hline

$\Delta$ log N/O      &   0.05 & $-0.27$ & $-0.12$ & $-0.03$ & 1.68 & 1.69 &    &    Marigo (2001) $^{ b}$ $Z =$ 0.004\\

\medskip
$\Delta$ log O/H     &    0.01 & 0.69 & 0.61 & 0.52 & 0.23 & 0.16 &    &    Marigo (2001) $^{b}$  $Z =$ 0.004\\

$\Delta$ log N/O      &   0.20   &  0.32   & 0.38 &  0.38 & 1.53  & 2.05 & 2.09  &    Karakas (2010) $^{b}$  $Z =$ 0.004\\
\medskip
$\Delta$ log O/H    &    0.00     &   $0.02$   &  $0.01$ &  $0.00$ & $0.00$  & $-0.19$ & $-0.40$  &   Karakas (2010) $^{b}$  $Z =$ 0.004\\

$\Delta$ log N/O      &		0.26 &	0.45 &	0.54 &	0.53 &	0.55 & &	0.69 & 	 Lagarde et al. (2012)		$^{d}$	$Z =$ 0.004\\		\medskip																														
$\Delta$ log O/H    &			0.01 &	0.00	 & $-0.02$ &	$-0.02$	 &-0.02 & &	$-0.02$ & 	Lagarde et al. (2012)		$^{d}$	$Z =$ 0.004\\	
					
$\Delta$ log N/O      &		0.51 &	0.55	& 0.63	& 0.66	& 0.82 & &	0.78	& 	Lagarde et al. (2012)		$^{e}$	$Z =$ 0.004\\	
			
$\Delta$ log O/H    &		    0.02 &	-0.01 & 	$-0.03$ &	$-0.03$ &	$-0.02$ & & 	0.02  & 	Lagarde et al. (2012)		$^{e}$	$Z =$ 0.004\\	
																																							
\hline
\multicolumn{7}{l}{$^{a}$ Non rotating solar metallicity models}\\
\multicolumn{7}{l}{$^{b}$  Synthetic  models, with $\alpha$=1.68 }\\
\multicolumn{7}{l}{$^{c}$ Detailed, post-processed  models }\\
\multicolumn{7}{l}{$^{d}$ Detailed  models; computations  until the end of the early AGB phase }\\
\multicolumn{7}{l}{$^{e}$ Detailed models with rotation; computations  until the end of the early AGB phase}\\
\end{tabular}
\end{table*}

We now try to understand the main features of the behavior of abundances in PNe and \hii\ regions in NGC 300 within a more general framework. The most outstanding findings requiring an explanation are:
\begin{enumerate}
  \item The N/O ratios in most PNe are larger by 0.4 -- 1 dex than  in  \hii\ regions at similar galactocentric distances.
  \item The O/H ratio for PNe presents a larger intrinsic scatter than for \hii\ regions\footnote{Part of the scatter observed in \hii\ regions at a given galactocentric distance is likely due to abundance uncertainties, as discussed in Sect. \ref{sec:gradients}}.
  \item The O/H radial gradient obtained from PNe is flatter than that obtained from \hii\ regions.
\end{enumerate}

\begin{figure}
\includegraphics[width=\columnwidth]{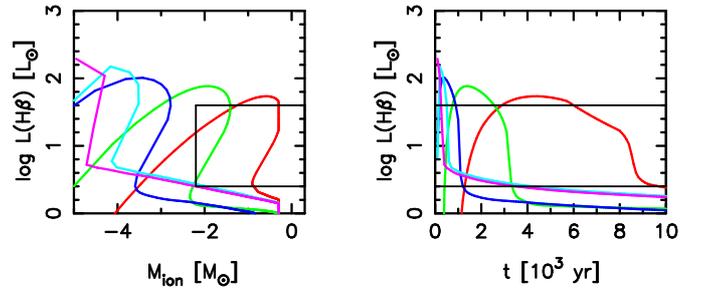} 
\caption{Evolution of the PN toy model described in Sect. \ref{sec:discussion} for PNe having progenitor  masses (from right to left) of 2\msun\ (red), 2.5\msun\ (green), 3\msun\ (blue), 4\msun\ (cyan), 5\msun\ (magenta). For  1\msun,  the \Hb\ luminosity would be too small to be detected. The black box in the left panel indicates the observed range for our PNe. The right panel indicates the time evolution of $L$(\Hb) since the formation of the PN model. It illustrates that the chance to observe a PN having a progenitor mass of 2\msun\ is much larger than for  2.5\msun. \label{fig:npevol}}
\end{figure}

In order to aid our discussion, we list in Table \ref{tab:pnprop} a few relevant properties of PN progenitors, i.e. of stars with initial masses  $M_{\rm i}$ between 1\,\msun\ and 7\,\msun. These are: the main sequence lifetimes, $t_{\rm MS}$, the ``final'' masses, i.e. the masses of the central stars of the PNe they produce, and some characteristics of the computed chemical yields: the values of $\Delta$ log N/O and $\Delta$ log O/H, which are the predicted changes of the PN chemical composition with respect to the initial composition of the progenitor atmosphere. Those values have been obtained from recent stellar evolutionary models (Marigo 2001, Karakas 2010 and Lagrange et al. 2012). They are listed for three different metallicities: solar, $Z = 0.008$ and $Z = 0.004$. In Fig. \ref{fig:npevol} we show the evolution of the PNe of different central star masses,  using  a simple toy model where a spherical nebula of homogeneous density and constant mass (0.5\,\msun) expands with a velocity of 20\kms around the central star. For the stellar evolution, we use the set of hydrogen-burning post-AGB evolutionary tracks of Bl\"ocker (1995, and references therein), with the same interpolation procedure as in Stasi\'nska et al. (1997). The stars are assumed to radiate like black bodies, and $L$(\Hb) has been computed analytically assuming an electron temperature of $10^{4}$~K. While obviously the evolution of real PNe is far more complex than that (for example the covering factor is not necessarily unity, the density is not homogeneous and the expansion velocity is not uniform), this simple toy model serves to fix order of magnitude properties. In the left panel, the black rectangle delimits the region covered by the PNe we observed in NGC 300. The right panel shows the time evolution of  $L$(\Hb) for the same models, starting from the ejection of the nebula. 

\begin{figure} [h]
\includegraphics[width=\columnwidth]{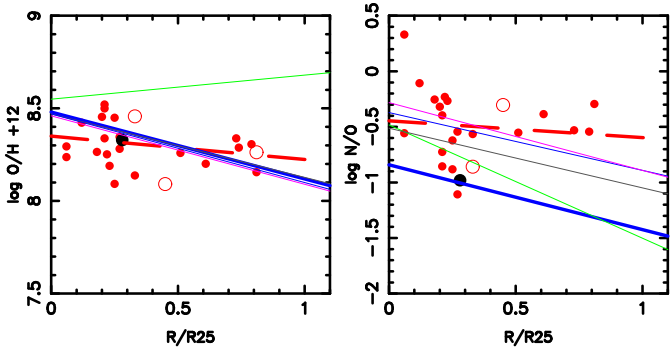} 
\caption{Observed and predicted radial trends of O/H and N/O in the PNe of NGC 300. The thin lines show the predicted trends assuming that the progenitor initial abundances follow the regression line for \hii\ regions (represented with the thick blue line). The different colors correspond to the different stellar evolution models listed in Table \ref{tab:pnprop}: green for Marigo (2001), grey for Karakas (2010), blue for the standard models of Lagarde et al. (2012), purple for the models with rotation of Lagarde et al. (2012). The thick dashed lines shows is regression line for the observed PN abundances.\label{fig:npyield}}
\end{figure}

By considering Fig. \ref{fig:npevol} and Table \ref{tab:pnprop}, one can see that the bulk of our  PNe likely arise from stars of initial masses around 2--2.5\,\msun, with masses closer to 2\,\msun\ being probably more frequent since their PNe are sufficiently bright for a longer time. This implies that the progenitors were born in a rather narrow time interval: roughly 0.5 - 1 Gyr ago, according to Table \ref{tab:pnprop}. 

It follows that the oxygen abundance scatter observed among PNe at any galactocentric distance is likely not due to sampling of different epochs of the galaxy chemical evolution. Another way to produce the observed abundance scatter would be the presence of radial stellar  motions. These would imply that a PN seen now at a certain galactocentric distance may have been formed in a radially distinct  zone. The importance of  radial stellar motions in NGC 300 has been assessed by Gogarten et al. (2010) by means of colour-magnitude diagrams of stellar populations in different radial bins. With the help of Monte Carlo simulations, these authors found that stars  in their model for NGC 300 do not undergo important radial migration.  This leaves us with  only one alternative to explain the O/H dispersion: the modification of the original abundances by nucleosynthesis and mixing during the progenitor evolution. Indeed, the abundance of oxygen  can be increased  by dredge up of He-burning products or decreased by hot bottom-burning. For example, the models of Marigo (2001) listed in Table \ref{tab:pnprop} show the effect of third dredge-up on oxygen in a stellar mass range appropriate to NGC 300. However, the models by Karakas (2010) as well as those by Lagarde et al. (2012), which include more complete physics than Marigo (2001), do not predict an important oxygen abundance variation. This is illustrated in Fig. \ref{fig:npyield}, which shows the radial variation of the oxygen abundance in the PNe of NGC 300, assuming that the initial abundance is equal to that of the \hii\ regions (represented by the thick blue line) and taking the predictions from the different models shown in Table  \ref{tab:pnprop} for an inital mass of 2.5\msun. Clearly, only the models of Marigo (2001) produce a noticeable trend. However,  this trend, which would reverse the gradient with respect to that of the \hii\ regions, is not in agreement with the observations. We should note, however, that the available grids of models were done using specific parametrization of processes such as mass-loss or convection and do not explore the entire parameter space.  One could think that, in different conditions, models could predict a different behavior of the oxygen abundance. For example, Decressin et al. (2009) find, for their solar metallicity models with initial rotation velocity of 300 km sec$^{-1}$ at the surface (i.e. larger than the Lagarde et al. models) that the oxygen abundance has decreased by 0.65 dex at the end of the early AGB phase.

We note that also in M 33 the scatter of the O/H values observed for PNe is  much larger than that observed for  \hii\ regions at similar galactocentric distances (Bresolin et al.~2010). Many O/H values observed in PNe in that galaxy are actually significantly higher than \textit{any} of the O/H values observed in \hii\ regions. This suggests that oxygen production in the PN progenitors in that galaxy is quite common, too.

Finally,  we have found that in NGC 300 the O/H radial gradient derived from PNe is flatter than that obtained from  \hii\  regions. Taken at face value, and ignoring the considerations above, this might be considered as evidence that the metallicity gradient in this galaxy is steepening with time, since PNe are an older population compared to \hii\ regions. However, we have just argued that the oxygen abundances in PNe are affected by nucleosynthesis in the progenitors and this can happen in both ways. Thus it is difficult to determine the metallicity gradient at birth from the observed O/H ratios.  We note, however, that argon  -- which is not suspected of being modified by progenitor nucleosynthesis -- also shows  a flatter galactic gradient in PNe than  in \hii\ regions (see Fig. \ref{fig:gradsH} and Table \ref{tab:resugrad}), although not so clearly as in the case of O/H, and the dispersion is even larger. However, as mentioned several times in this paper, the ICFs for argon are very crude, so the argon abundances are not fully reliable. On the other hand, there is a priori no reason why uncertainties in the argon abundances would conspire to produce an artificial flattening of the gradient. The available data on PNe would therefore indicate that there is a steepening of the metallicity gradient in NGC 300 over the last Gyr, however the evidence is not really strong. One worry is  the fact that at $R/R_{25}$\,=\,0.8, the argon abundances seen in PNe are larger than those in the interstellar medium as traced by the \hii\ regions by about 0.1--0.2 dex. It is not clear whether this is significant. Sulfur is another element that is not suspected of being modified by progenitor nucleosynthesis. However, as mentioned in Sect. \ref{sec:uncertainties}, there is a well documented problem with sulfur abundances in PNe, so until this problem is not solved, it is not recommended to use PN sulfur abundances to test chemical evolution scenarios.

Since we have argued that the oxygen abundances of the PNe observed in NGC 300 have been affected by nucleosynthesis in their progenitors, there is one more topic to be discussed. This is the small dispersion in the Ne/O ratio in PNe and the fact that this ratio is is higher than the one in \hii\ regions by about 0.2 dex. It is known that, while $^{20}$Ne, the main isotope of neon, does not change its abundance during the evolution of intermediate mass stars, $^{22}$Ne can be produced through combined H and He burning during the thermally pulsing evolution (Karakas \& Lattanzio 2003). The range of stellar masses in which this production can significantly affect the total Ne abundance in the models of Karakas (2010) is $M_{\rm i} = $2.5 -- 3.5 \msun\ at a metallicity $Z=0.008$  and $M_{\rm i}  > 2\,$\msun\ at a metallicity $Z=0.004$. So the PNe of our sample could well be affected by this process, which would explain the fact that their Ne/O ratio is larger than in \hii\ regions. The question is: why are the O and Ne abundances so strongly correlated, since they are not produced by the same process, as discussed in Karakas \& Lattanzio (2003)? (Note that we have argued in Sect.  \ref{sec:patterns} that the ICFs for neon might be inadequate, but from Fig. \ref{fig:icfs} the Ne/O ratios calculated for our PNe do not depend on excitation).

 As regards the behavior of the N/O ratio in the PNe,  Table \ref{tab:pnprop} shows  that for such progenitor masses as relevant for our NGC 300 sample,  PNe are expected to show a N/O enhancement of about 0.3 to 0.6 dex (depending on model and metalicity).  Our observations, however, show that, in the central parts of the Galaxy, N/O ratios in PNe range from -1.2 (i.e. close to the value found in \hii\ regions) to 0.35. So the observed dispersion is much larger than predicted. In the outer part of the galaxy, they are all above -0.4. The thin lines representing the radial behavior for N/O for a progenitor mass of 2.5 \msun\ do not follow the observed behavior very well, for any of the considered models. As in the case of the O/H ratio, the observed behavior of the N/O ratio in PNe argues for a large variety of conditions that affect the mixing processes. Note that the behavior of N/O  in NGC 300 is actually similar that in M 33 (Bresolin et al. 2010). This means that the requirements for stellar evolution models are similar for both galaxies, which is reassuring.

\section{Summary}
\label{summary}

We have obtained high signal-to-noise spectroscopy with the VLT of 26 PNe and 9 compact \hii\ regions in the nearby spiral galaxy NGC 300. These objects were detected in the \oiii\ on- and off-band imaging survey of Pe\~na et al. (2012) and supplement the data obtained by Bresolin et al. (2009) for a sample of giant \hii\ regions. 
After a discussion of the available atomic data, we have determined the abundances of He, N, O, Ne, S and Ar in all these objects in a consistent way. 

We have discussed the nature of the candidate PNe and compact \hii\ regions in the light of our new spectroscopic information. It turns out that one object that was identified as a candidate PN from the photometric data is in fact a compact \hii\ region, while one candidate compact \hii\ region is in fact a PN. 

Concerning the general abundance patterns shown in Fig. \ref{fig:abpatterns}, we find, not surprisingly, that compact \hii\ regions have chemical abundances that are globally similar to those of giant \hii\ regions. PNe have systematically larger N/O ratios compared to \hii\ regions, while they have similar Ne/O and Ar/O ratios. The sulfur abundances in PNe show an unphysical trend with excitation, probably indicating an inadequate ionization correction factor for this element. All theses results are in rough agreement with recent findings in other galaxies such as the Small and Large Magellanic Clouds (Leisy \& Dennefeld 2006), NGC 3109 (Pe\~na et al.~2007), NGC 6822 (Hern{\'a}ndez-Mart{\'{\i}}nez et al.~2009), and M33 (Bresolin et al. 2011).
The large N/O ratios in PNe as compared to \hii\ regions is an indication that N has been altered by nucleosynthesis in the progenitors. While similar results obtained in other galaxies are generally interpreted as being the result of second dredge-up, this cannot be the case here, since we have shown that most of the PNe in our sample likely arise from stars of initial masses around 2--2.5\,\msun.  An extra mixing process is required to bring the freshly produced N to the stellar surface before the ejection of the PN, likely driven by rotation.

With regard to the radial behavior of the chemical abundances across the galaxy, we find that compact \hii\ regions follow the giant \hii\ regions rather well (with however a tendency for compact \hii\ to have slightly lower abundances with respect to hydrogen than giant \hii\ rgions). On the other hand,  PNe behave very differently. In the central parts of the galaxy PNe have an average O/H abundance ratio 0.15 dex smaller than \hii\ regions.   The abundance dispersion at any galactocentric radius is significantly larger for PNe than for \hii\ regions. This suggests that the oxygen abundance in PNe is also affected by nucleosynthesis in their progenitors, in a way which depends on several parameters.   The formal O/H, Ne/H and Ar/He abundance gradients for PNe are shallower than for  \hii\ regions. In fact, they are almost flat. We have argued that this indicates a steepening of the metallicity gradient in NGC 300 over the last Gyr, rather than the effect of radial stellar motions. However the large observed abundance dispersion  makes any conclusion on a possible steepening of the gradients with time only tentative.

We end this study by recalling several problems that are general to the study of the chemical abundances of PNe, which we have not considered here and should be kept in mind. One is the unsatisfactory situation with the ionization correction factors used to compute the abundances: this is known to be important for the sulfur abundance, but all the other elements, including oxygen, might be affected to a certain degree. An additional problem is the amount of depletion into dust grains, which has been brought up by Rodr{\'{\i}}guez \&  Delgado-Inglada (2011) for  PNe and \hii\ regions in the solar vicinity. Finally, there is the still unsolved problem of the discrepancy between abundances derived from collisionally excited and from recombination lines (see e.g. the book by Stasi\'nska et al.~2012 for a summary), which casts some doubt on the relevance of the computed abundances. The reliability of the conclusions to which we arrived at in this study -- as in any other study of this kind -- depends somehow on the solutions that will be found to those problems.

\begin{acknowledgements}
M. P. is grateful to DAS, Universidad de Chile, for hospitality during a sabbatical stay when part of this work was done. 
  This work received financial support from grant  IN-105511(PAPIIT DGAPA-UNAM).
FB gratefully acknowledges partial support from the National Science Foundation
grants AST-0707911 and AST-1008798.
YGT acknowledges the award of a Marie Curie intra-European
Fellowship within the 7th European Community Framework Programme
(grant agreement PIEF-GA-2009-236486).
\end{acknowledgements}

\tiny

\clearpage

Table 3 for the on-line edition

Table 3a
\begin{verbatim}

# imag-list              2                 5                 8                12                14                20                22
R.A.(h m s)          00 54 33.17       00 54 35.39       00 54 36.05       00 54 37.89       00 54 38.91       00 54 41.59       00 54 42.23
Dec (o ' ")          -37 38 26.7       -37 39 36.0       -37 39 50.8       -37 40 14.0       -37 39 43.2       -37 40 21.4       -37 40 04.8
R/R25                 0.49              0.40              0.38              0.33              0.33              0.25              0.25          
                      CHR               CHR               CHR               PN                PN                PN                PN          
lambda_0     ion      I/I(Hb)  err      I/I(Hb)  err      I/I(Hb)  err      I/I(Hb)  err      I/I(Hb)  err      I/I(Hb)  err      I/I(Hb)  err
3726+29      [OII]    4.504   0.200     2.485   0.040     2.882   0.200     0.608   0.150     0.643   0.050   < 0.870           < 0.700
3750.15        H12                      0.027   0.020     0.037   0.001                                                                                 
3770.63        H11    0.043   0.015     0.037   0.020     0.032   0.007                                                                       
3797.90        H10    0.042   0.015     0.042   0.020     0.030   0.006                                                                       
3835.39         H9    0.056   0.012     0.059   0.020     0.052   0.010                       0.078   0.039                                   
3868.75    [NeIII]    0.048   0.013     0.115   0.012     0.042   0.004     0.674   0.168     0.796   0.040                       0.982   0.060
3889.05     HeI+H8    0.182   0.020     0.162   0.015     0.169   0.013   < 0.100             0.165   0.033                       0.388   0.050
3970.07 [NeIII]+H7    0.169   0.017     0.177   0.015     0.148   0.012   < 0.100             0.450   0.045                                     
4026.21        HeI    0.036   0.016     0.017   0.008     0.013   0.007   < 0.100             0.043   0.022                                     
4068.60      [SII]    0.026   0.013     0.022   0.010     0.030   0.015   < 0.100             0.061   0.031                                    
4101.74         Hd    0.255   0.020     0.260   0.020     0.248   0.012     0.196   0.098     0.256   0.015     0.255   0.033     0.245   0.030
4267.15        CII                                                        < 0.100           < 0.050                             < 0.100          
4340.47         Hg    0.471   0.020     0.470   0.020     0.468   0.023     0.486   0.050     0.478   0.020     0.474   0.033     0.473   0.024
4363.21     [OIII]    0.015   0.006     0.011   0.003     0.008   0.004     0.172   0.060     0.212   0.020   < 0.075             0.250   0.038
4471.47        HeI    0.043   0.005     0.036   0.004     0.033   0.004   < 0.100             0.044   0.022                                     
4685.68       HeII                                                          0.632   0.095     0.245   0.015   < 0.027           < 0.080           
4711.37     [ArIV]                                                        < 0.100           < 0.030                                            
4740.17     [ArIV]                                                        < 0.100             0.078   0.039                                      
4861.33         Hb    1.000   0.015     1.000   0.002     1.000   0.020     1.000   0.060     1.000   0.030     1.000   0.040     1.000   0.020
4921.93        HeI    0.008   0.004     0.010   0.005     0.012   0.006                                                                              
4958.91     [OIII]    0.529   0.020     0.754   0.015     0.437   0.020     4.700   0.280     3.325   0.033     2.314   0.070     3.940   0.039
5006.84     [OIII]    1.525   0.015     2.214   0.040     1.308   0.020    14.380   0.860     9.690   0.097     6.772   0.200    11.400   0.114
5015.68        HeI    0.021   0.003     0.016   0.006     0.013   0.006                                                                                 
5200.26       [NI]                                                                                                                                
5754.68      [NII]                                                                                                                             
5875.64        HeI    0.130   0.018     0.373   0.009     0.105   0.010                       0.148   0.013     0.086   0.009     0.138   0.030
6300.30       [OI]    0.029   0.006     0.011   0.002     0.016   0.008                       0.057   0.008                     < 0.100         
6312.12     [SIII]    0.012   0.006     0.016   0.002     0.011   0.005                       0.022   0.005                                       
6548.03      [NII]    0.112   0.006     0.099   0.010     0.106   0.015                       0.174   0.015     0.104   0.010   < 0.100         
6562.82         Ha    2.859   0.020     2.861   0.040     2.861   0.020     2.860   0.080     2.855   0.020     2.860   0.080     2.870   0.030
6583.41      [NII]    0.334   0.017     0.296   0.014     0.316   0.015     0.160   0.080     0.526   0.010     0.283   0.023     0.275   0.050
6678.15        HeI    0.032   0.004     0.035   0.005     0.028   0.004                       0.033   0.006     0.035   0.007     0.043   0.030
6716.47      [SII]    0.244   0.001     0.173   0.013     0.181   0.012                       0.041   0.012     0.118   0.010     0.108   0.020
6730.81      [SII]    0.177   0.001     0.133   0.012     0.126   0.012                       0.085   0.007     0.120   0.010     0.095   0.020
7065.28        HeI    0.014   0.004     0.021   0.003     0.016   0.002                       0.049   0.009     0.106   0.090     0.054   0.030
7135.78    [ArIII]    0.085   0.007     0.097   0.012     0.086   0.008                       0.151   0.008     0.186   0.015     0.150   0.015
7319         [OII]    0.038   0.004     0.020   0.003     0.014   0.004                       0.100   0.010     0.042   0.007   + 0.093   0.040
7330         [OII]    0.026   0.003     0.018   0.003     0.019   0.004                       0.055   0.020     0.077   0.011                  
9068.60     [SIII]    0.238   0.020     0.325   0.040     0.285   0.029                       0.350   0.070     0.533   0.060     0.169   0.015
9529.78     [SIII]    0.598   0.040     0.487   0.040     0.761   0.030                       0.785   0.150     1.207   0.120     0.245   0.021
                                                                                                                                           
c(Hb)                 0.15              0.00              0.12              0.00              0.00              1.17              0.44         
log F(Hb)             -15.103           -14.524           -14.555          -16.590            -15.920           -16.350           -16.080   


\end{verbatim}
\clearpage

Table 3b
\begin{verbatim}


# imag-list             24                25                35                39                40                45                48
R.A.(h m s)          00 54 43.70       00 54 44.42       00 54 48.38       00 54 51.25       00 54 52.08       00 54 53.82       00 54 54.91
Dec (o ' ")          -37 41 51.3       -37 41 29.4       -37 39 48.4       -37 41 46.2       -37 42 43.2       -37 39 27.5       -37 41 32.4
R/R25                 0.25              0.21              0.18              0.11              0.22              0.21              0.06
                      PN                PN                PN                CHR               PN                PN                PN 
lambda_0       ion    I/I(Hb)  err      I/I(Hb)  err      I/I(Hb)  err      I/I(Hb)  err      I/I(Hb)  err      I/I(Hb)  err      I/I(Hb)  err
3726+29      [OII]    0.466   0.070     1.038   0.120     0.584   0.060     1.168   0.150     0.222   0.060     0.592   0.180     0.350   0.150
3750.15        H12                                                                                                                                                                                                                                     
3770.63        H11                                                                                                                                    
3797.90        H10                                                                                                                              
3835.39         H9    0.119   0.060   < 0.100             0.059   0.020     0.059   0.008     0.055   0.025     0.085   0.060   < 0.100        
3868.75    [NeIII]    1.010   0.050     0.980   0.070     0.980   0.060     0.131   0.019     0.515   0.050     0.533   0.060     0.941   0.080
3889.05     HeI+H8    0.179   0.015     0.148   0.050     0.179   0.030     0.157   0.022     0.139   0.050     0.188   0.040     0.216   0.070
3970.07 [NeIII]+H7    0.365   0.040     0.537   0.050     0.600   0.030     0.229   0.034     0.339   0.050     0.362   0.060     0.452   0.090
4026.21        HeI  < 0.080           < 0.070             0.030   0.020     0.014   0.007   < 0.050           < 0.040           < 0.090          
4068.60      [SII]  < 0.080           < 0.070             0.041   0.020                     < 0.050           < 0.040           < 0.090         
4101.74         Hd    0.290   0.040     0.275   0.030     0.302   0.040     0.278   0.028     0.244   0.030     0.312   0.060     0.247   0.070
4267.15        CII  < 0.080           < 0.070             0.035   0.020                     < 0.050           < 0.050           < 0.090          
4340.47         Hg    0.470   0.020     0.468   0.030     0.465   0.030     0.468   0.037     0.482   0.030     0.462   0.030     0.468   0.030
4363.21     [OIII]    0.226   0.040     0.176   0.025     0.245   0.030     0.015   0.006     0.113   0.025     0.051   0.025     0.219   0.050
4471.47        HeI  < 0.100           < 0.050             0.077   0.020     0.024   0.007     0.050   0.030     0.045   0.030   < 0.100         
4685.68       HeII    0.469   0.020     0.290   0.020     0.163   0.019                       0.031   0.020   < 0.080           < 0.100        
4711.37     [ArIV]    0.053   0.020     0.028   0.009     0.046   0.007                     < 0.030           < 0.080           < 0.100        
4740.17     [ArIV]    0.079   0.020     0.066   0.009     0.089   0.010                     < 0.030           < 0.080           < 0.100          
4861.33         Hb    1.000   0.020     1.000   0.020     1.000   0.001     1.000   0.020     1.000   0.010     1.000   0.010     1.000   0.020
4921.93        HeI                                                                                                                               
4958.91     [OIII]    4.536   0.020     4.199   0.040     4.366   0.130     0.928   0.015     3.215   0.090     2.868   0.030     4.514   0.080
5006.84     [OIII]   13.518   0.020    12.431   0.040    12.790   0.250     2.840   0.040     9.212   0.300     8.062   0.080    13.860   0.300
5015.68        HeI                                                                                                                             
5200.26       [NI]                                                                                                                         
5754.68      [NII]                                                                                                                0.108   0.030
5875.64        HeI    0.091   0.030     0.187   0.022     0.197   0.017     0.073   0.010     0.138   0.046     0.169   0.030     0.200   0.040
6300.30       [OI]                      0.114   0.021     0.095   0.019   < 0.070           < 0.100           < 0.100             0.083   0.020
6312.12     [SIII]                      0.037   0.007   < 0.040           < 0.070           < 0.100           < 0.070             0.051   0.020
6548.03      [NII]    0.114   0.015     0.266   0.020     0.315   0.012     0.182   0.020     0.098   0.024     0.074   0.040     0.408   0.020
6562.82         Ha    2.860   0.050     2.860   0.040     2.859   0.050     2.862   0.050     2.860   0.050     2.858   0.050     2.860   0.020
6583.41      [NII]    0.276   0.015     0.836   0.038     0.992   0.030     0.571   0.050     0.246   0.007     0.204   0.030     1.120   0.040
6678.15        HeI  < 0.080             0.026   0.012     0.059   0.025     0.018   0.009     0.042   0.008     0.041   0.030     0.052   0.030
6716.47      [SII]  < 0.080           < 0.046             0.034   0.011     0.348   0.020   < 0.040             0.059   0.030     0.078   0.030
6730.81      [SII]  < 0.080             0.076   0.015     0.061   0.006     0.245   0.020   < 0.040             0.085   0.030     0.105   0.030
7065.28        HeI  < 0.100             0.104   0.010     0.108   0.013     0.023   0.007     0.106   0.003     0.099   0.035     0.070   0.010
7135.78    [ArIII]    0.176   0.022     0.180   0.015     0.140   0.014     0.043   0.006     0.095   0.004     0.056   0.030     0.227   0.020
7319         [OII]  < 0.100             0.115   0.012     0.071   0.021     0.053   0.006     0.151   0.024   < 0.050             0.074   0.018
7330         [OII]                      0.075   0.010     0.092   0.018                       0.089   0.010   < 0.050             0.134   0.020
9068.60     [SIII]    0.529   0.030     0.260   0.040     0.186   0.037     0.139   0.002     0.207   0.040     0.103   0.040     0.271   0.050
9529.78     [SIII]    0.767   0.040     0.530   0.060     0.375   0.075     0.301   0.003     0.395   0.080   < 0.247             0.620   0.120
                                                                                                                                                
c(Hb)                 0.12              0.07              0.00              0.06              0.00              0.00              0.21           
log F(Hb)            -16.200            -16.050           -15.800           -15.349           -16.020           -16.050           -16.230      

\end{verbatim}
\clearpage

Table 3c
\begin{verbatim}


# imag-list             51                54                57                58                63                65                66
R.A.(h m s)           00 54 55.33       00 54 56.83       00 54 57.42       00 54 58.12       00 54 59.72       00 55 01.71       00 55 02.44
Dec (o ' ")           -37 41 28.5       -37 39 43.5       -37 41 01.0       -37 40 44.9       -37 39 26.1       -37 40 29.4       -37 39 54.6
R/R25                 0.06              0.20              0.09              0.12              0.27              0.21              0.27         
                      PN                PN                CHR               PN                PN                PN                PN          
lambda_0       ion    I/I(Hb)  err      I/I(Hb)  err      I/I(Hb)  err      I/I(Hb)  err      I/I(Hb)  err      I/I(Hb)  err      I/I(Hb)  err
3726+29      [OII]    0.595   0.028     2.620   0.080     3.279   0.013     2.207   0.050     1.709   0.080     0.138   0.060     0.639   0.038
3750.15        H12                                        0.037   0.025                                                                        
3770.63        H11                                        0.057   0.028                                                                        
3797.90        H10                                        0.044   0.015                                                                       
3835.39         H9    0.070   0.025   < 0.080             0.051   0.015                       0.070   0.035     0.078   0.040     0.063   0.031
3868.75    [NeIII]    0.479   0.029     0.960   0.060     0.498   0.050     1.194   0.070     0.135   0.060     0.743   0.030     0.736   0.030
3889.05     HeI+H8    0.166   0.042     0.114   0.034     0.173   0.020     0.110   0.030     0.124   0.040     0.255   0.076     0.173   0.020
3970.07 [NeIII]+H7    0.281   0.020     0.350   0.019     0.281   0.030     0.530   0.070     0.178   0.040     0.438   0.040     0.341   0.034
4026.21        HeI                    < 0.050             0.038   0.018                                         0.137   0.030   < 0.080        
4068.60      [SII]                    < 0.050             0.057   0.020     0.091   0.040                     < 0.140           < 0.060        
4101.74         Hd    0.232   0.018     0.203   0.018     0.271   0.020     0.252   0.030     0.232   0.030     0.325   0.050     0.244   0.030
4267.15        CII                    < 0.050                               0.034   0.015                       0.058   0.030   < 0.060   
4340.47         Hg    0.474   0.015     0.490   0.013     0.468   0.030     0.467   0.017     0.467   0.046     0.460   0.023     0.469   0.015
4363.21     [OIII]    0.089   0.014     0.104   0.012     0.026   0.007     0.208   0.017     0.023   0.010     0.146   0.015     0.149   0.013
4471.47        HeI    0.078   0.012     0.074   0.010     0.048   0.009     0.044   0.030     0.067   0.010     0.033   0.040     0.077   0.015
4685.68       HeII  < 0.064             0.044   0.020     0.024   0.006     0.352   0.017   < 0.011             0.159   0.016     0.148   0.022
4711.37     [ArIV]                    < 0.040                             < 0.040                                               < 0.040        
4740.17     [ArIV]                    < 0.040                             < 0.040                                               < 0.040        
4861.33         Hb    1.000   0.014     1.000   0.019     1.000   0.020     1.000   0.017     1.000   0.050     1.000   0.030     1.000   0.020
4921.93        HeI                                        0.013   0.005                                                                        
4958.91     [OIII]    2.668   0.038     3.255   0.062     1.603   0.030     4.371   0.076     1.152   0.050     3.957   0.119     3.768   0.040
5006.84     [OIII]    7.849   0.110     9.592   0.182     4.780   0.090    13.314   0.240     3.470   0.150    11.250   0.338    11.424   0.114
5015.68        HeI                                                                                                                                   
5200.26       [NI]                                        0.025   0.008                                                                                                                  
5754.68      [NII]                      0.069   0.035                       0.063   0.020                                                     
5875.64        HeI    0.174   0.021     0.146   0.012     0.125   0.012     0.120   0.016     0.127   0.013     0.132   0.016     0.138   0.012
6300.30       [OI]  < 0.012             0.197   0.014     0.128   0.013     0.171   0.021     0.017   0.005   < 0.080             0.154   0.014
6312.12     [SIII]  < 0.012             0.042   0.012     0.029   0.005     0.050             0.016   0.005   < 0.080           < 0.060       
6548.03      [NII]    0.123   0.007     0.846   0.030     0.270   0.027   < 0.776   0.017     0.095   0.009   < 0.080             0.115   0.015
6562.82         Ha    2.860             2.860   0.030     2.861   0.050     2.860   0.030     2.860   0.100     2.860   0.100     2.860   0.020
6583.41      [NII]    0.323   0.002     2.580   0.030     0.799   0.070     2.360   0.023     0.270   0.021     0.101   0.050     0.228   0.028
6678.15        HeI    0.037   0.010     0.053   0.020     0.039   0.010     0.052   0.026     0.032   0.005   < 0.080             0.042   0.008
6716.47      [SII]  < 0.030             0.152   0.015     0.649   0.060     0.222   0.015     0.090   0.008   < 0.080             0.060   0.007
6730.81      [SII]  < 0.030             0.238   0.025     0.461   0.050     0.259   0.010     0.099   0.008   < 0.080             0.046   0.010
7065.28        HeI    0.070   0.009     0.074   0.009     0.022   0.050     0.068   0.011     0.046   0.005   < 0.080             0.075   0.008
7135.78    [ArIII]    0.142   0.015     0.233   0.030     0.185   0.020     0.237   0.014     0.131   0.013   < 0.080             0.082   0.012
7319         [OII]  + 0.105   0.020     0.141   0.014     0.036   0.004     0.084   0.011     0.035   0.006                     < 0.080       
7330         [OII]                      0.128   0.030     0.027   0.004     0.100   0.030     0.027   0.010                       0.114   0.014
9068.60     [SIII]    0.227   0.030     0.362   0.070     0.340   0.030     0.326   0.065     0.466   0.060   < 0.250             0.106   0.210
9529.78     [SIII]    0.488   0.050     0.742   0.100     0.807   0.080     0.824   0.100     1.183   0.170   < 0.250             0.317   0.063
                                                                                                                                          
c(Hb)                 0.00              0.00              0.09              0.00              0.41              0.00              0.00         
log F(Hb)             -15.920           -16.100           -15.113           -16.040           -15.733           -16.310           -16.110        

\end{verbatim}
\clearpage

Table 3d
\begin{verbatim}


# imag-list             69                74                87                88                90a               90b               91
R.A.(h m s)           00 55 03.97       00 55 05.77       00 55 13.77       00 55 13.78       00 55 15.08       00 55 15.08       00 55 15.91
Dec (o ' ")           -37 40 53.3       -37 42 11.9       -37 41 39.2      -37 40 32.7        -37 44 14.5       -37 44 14.5       -37 43 20.4
R/R25                 0.23              0.28              0.43              0.45              0.55              0.55              0.51
                      PN (S6)           PN                CHR               PN                CHR               CHR               PN
lambda_0       ion    I/I(Hb)  err      I/I(Hb)  err      I/I(Hb)  err      I/I(Hb)  err      I/I(Hb)  err      I/I(Hb)  err      I/I(Hb)  err
3726+29      [OII]    0.284   0.070     1.150   0.040     3.244   0.100     1.158   0.080     2.919   0.140     2.998   0.150     0.698   0.070
3750.15        H12                                        0.040   0.010                       0.038   0.006     0.038   0.011                         
3770.63        H11                                        0.050   0.010                       0.048   0.005     0.047   0.011                     
3797.90        H10                                        0.054   0.010                       0.049   0.005     0.051   0.011                    
3835.39         H9                      0.043   0.030     0.071   0.015                       0.068   0.007     0.076   0.014     0.080   0.020
3868.75    [NeIII]    0.686   0.062     0.266   0.030     0.093   0.015     0.658   0.100     0.123   0.010     0.111   0.013     0.660   0.040
3889.05     HeI+H8    0.144   0.022     0.092   0.030     0.146   0.015     0.248   0.070     0.182   0.015     0.185   0.016     0.174   0.020
3970.07 [NeIII]+H7    0.382   0.038     0.192   0.030     0.164   0.015     0.412   0.060     0.189   0.015     0.170   0.015     0.311   0.031
4026.21        HeI  < 0.090                               0.011   0.006                       0.018   0.005     0.027   0.014                 
4068.60      [SII]  < 0.090                               0.013   0.006                       0.016   0.005     0.021   0.014                 
4101.74         Hd    0.244   0.020     0.227   0.030     0.244   0.020     0.314   0.050     0.254   0.020     0.254   0.020     0.243   0.020
4267.15        CII                                                                                                                            
4340.47         Hg    0.478   0.020     0.466   0.030     0.468   0.020     0.469   0.035     0.467   0.030     0.468   0.020     0.469   0.015
4363.21     [OIII]    0.124   0.020   : 0.034   0.015     0.019   0.004     0.150   0.030     0.020   0.005     0.017   0.005     0.227   0.015
4471.47        HeI    0.091   0.020     0.039   0.015     0.036   0.004                       0.043   0.004     0.042   0.005                
4685.68       HeII  < 0.040           < 0.020                               0.492   0.040                                         0.466   0.020
4711.37     [ArIV]  < 0.040                                                                                                       0.079   0.014
4740.17     [ArIV]  < 0.040                                                                                                       0.070   0.013
4861.33         Hb    1.000   0.020     1.000   0.020     1.000   0.020     1.000   0.040     1.000   0.020     1.000   0.002     1.000   0.020
4921.93        HeI                                        0.011   0.005                       0.012   0.006     0.013   0.003                    
4958.91     [OIII]    3.036   0.030     1.693   0.030     0.689   0.020     3.149   0.150     0.871   0.020     0.820   0.030     4.230   0.040
5006.84     [OIII]    9.160   0.090     4.937   0.100     2.029   0.020     8.762   0.200     2.548   0.050     2.419   0.020    11.822   0.100
5015.68        HeI                                        0.024   0.007                       0.027   0.010     0.022   0.005                   
5200.26       [NI]                                        0.018   0.005                       0.006   0.002     0.024   0.006                    
5754.68      [NII]                                                                                                                                  
5875.64        HeI    0.151   0.013     0.154   0.015     0.117   0.010   < 0.170             0.133   0.009     0.140   0.014     0.120   0.020
6300.30       [OI]    0.046   0.009                       0.024   0.004                       0.015   0.005                                    
6312.12     [SIII]  < 0.040                               0.025   0.004                       0.020   0.006                                   
6548.03      [NII]    0.095   0.007     0.052   0.010     0.123   0.018     0.206   0.040     0.094   0.009     0.081   0.008     0.103   0.026
6562.82         Ha    2.860   0.020     2.860   0.040     2.860   0.040     2.860   0.050     2.860   0.030     2.860   0.030     2.860   0.060
6583.41      [NII]    0.235   0.018     0.232   0.020     0.363   0.020     0.688   0.080     0.268   0.024     0.207   0.010     0.256   0.015
6678.15        HeI    0.051   0.020     0.035   0.040     0.030   0.003   < 0.070             0.037   0.010     0.044   0.010                     
6716.47      [SII]    0.040   0.019     0.080   0.010     0.171   0.013     0.396   0.060     0.141   0.012     0.072   0.007     0.029   0.006
6730.81      [SII]    0.050   0.010     0.081   0.010     0.136   0.013     0.385   0.070     0.120   0.012     0.086   0.008     0.058   0.010
7065.28        HeI    0.098   0.006     0.080   0.008     0.030   0.008                       0.028   0.008     0.066   0.007     0.031   0.008
7135.78    [ArIII]    0.098   0.007     0.172   0.010     0.092   0.013     0.172   0.020     0.112   0.010     0.124   0.008     0.102   0.015
7319         [OII]    0.112   0.007     0.052   0.010     0.032   0.003                       0.034   0.004     0.055   0.005                  
7330         [OII]    0.084   0.007     0.043   0.010     0.026   0.003                       0.030   0.005     0.042   0.005                   
9068.60     [SIII]    0.173   0.009     0.410   0.040     0.300   0.030   < 0.700             0.344   0.034     0.234   0.024     0.165   0.030
9529.78     [SIII]    0.339   0.017     1.081   0.050     0.896   0.030   < 0.700             0.844   0.044     0.490   0.024     0.586   0.170
                                                                                                                                       
c(Hb)                 0.00              0.64              0.36              0.00              0.36              0.35              0.00
log F(Hb)             -15.970           -15.510           -14.682           -16.570           -14.962           -15.043           -16.230

\end{verbatim}
\clearpage

Table 3e
\begin{verbatim}


# imag-list             96                97               102               103               104               108               109
R.A.(h m s)           00 55 22.54       00 55 23.67       00 55 27.20       00 55 27.49       00 55 28.86       00 55 30.30       00 55 30.54
Dec (o ' ")           -37 42 16.6       -37 44 55.9       -37 43 43.6       -37 40 55.1       -37 44 40.1       -37 41:15.7       -37 43 59.8
R/R25                 0.61              0.74              0.74              0.73              0.81              0.79              0.81
                      PN                PN                CHR               PN                PN                PN                PN
lamda_0        ion    I/I(Hb)  err      I/I(Hb)  err      I/I(Hb)  err      I/I(Hb)  err      I/I(Hb)  err      I/I(Hb)  err      I/I(Hb)  err  
3726+29      [OII]    0.223   0.045     noisy             3.544   0.200     0.433   0.043     1.250   0.075      ---              ---
3750.15        H12                                        0.026   0.013                                                                      
3770.63        H11                                                                                              
3797.90        H10                                        0.030   0.013                                                   
3835.39         H9                                        0.041   0.013     0.071   0.030     0.059   0.020     0.084   0.016     ---
3868.75    [NeIII]    0.777   0.080     0.623   0.120     0.076   0.013     0.986   0.050     0.819   0.050     0.701   0.040     0.854   0.060
3889.05     HeI+H8    0.206   0.030     0.179   0.080     0.170   0.017     0.178   0.018     0.157   0.015     0.152   0.015     0.183   0.030
3970.07 [NeIII]+H7    0.439   0.040     0.354   0.060     0.132   0.017     0.515   0.030     0.427   0.030     0.401   0.040     0.394   0.060
4026.21        HeI                                                          0.046   0.020                                                 
4068.60      [SII]                                                          0.037   0.020                                                    
4101.74         Hd    0.242   0.024     0.260   0.020     0.230   0.025     0.274   0.027     0.256   0.020     0.267   0.016     0.263   0.026
4267.15        CII                                                          0.015   0.007                                                     
4340.47         Hg    0.470   0.030     0.480   0.040     0.474   0.050     0.468   0.025     0.469   0.028     0.469   0.028     0.469   0.030
4363.21     [OIII]    0.178   0.020     0.110   0.040     0.017   0.005     0.199   0.020     0.170   0.020     0.165   0.025     0.130   0.020
4471.47        HeI                                        0.034   0.004     0.061   0.030                       0.055   0.020     0.062   0.030
4685.68       HeII    0.200   0.020   < 0.100                               0.144   0.012     0.550   0.020     0.119   0.012     0.205   0.030
4711.37     [ArIV]                    < 0.060                               0.036   0.020                       0.033   0.020     0.024   0.012
4740.17     [ArIV]                    < 0.060                               0.061   0.030                       0.031   0.020     0.030   0.015
4861.33         Hb    1.000   0.020     1.000   0.050     1.000   0.050     1.000   0.020     1.000   0.020     1.000   0.020     1.000   0.030
4921.93        HeI                                                                                                                           
4958.91     [OIII]    3.582   0.070     3.324   0.100     0.630   0.050     4.356   0.090     2.980   0.060     3.773   0.070     3.607   0.060
5006.84     [OIII]   10.104   0.200     9.700   0.290     1.840   0.080    12.796   0.026     8.780   0.100    11.328   0.230    10.229   0.200
5015.68        HeI                                        0.021   0.006                                     
5200.26       [NI]                                                                                                               
5754.68      [NII]                                                                          
5875.64        HeI  : 0.069             0.156   0.020     0.108   0.010     0.195   0.023     0.141   0.020     0.152   0.015     ---
6300.30       [OI]                                                          0.087   0.040                                       
6312.12     [SIII]                                        0.014   0.004     0.027   0.010                                      
6548.03      [NII]  < 0.080           < 0.100             0.080   0.008     0.137   0.010     0.290   0.050     0.101   0.020     ---   
6562.82         Ha    2.860   0.140     2.860   0.050     2.860   0.090     2.860   0.080     2.860   0.050     2.860   0.100     ---    
6583.41      [NII]    0.143   0.040   < 0.180             0.240   0.025     0.374   0.010     0.802   0.060     0.300   0.030     ---
6678.15        HeI                    < 0.180             0.034   0.007     0.046   0.010   < 0.040             0.041   0.030     ---
6716.47      [SII]  < 0.080           < 0.200             0.228   0.016     0.015   0.002     0.108   0.030     0.038   0.012     ---
6730.81      [SII]  < 0.080           < 0.200             0.161   0.009     0.039   0.010     0.131   0.030     0.073   0.014     ---
7065.28        HeI  < 0.080             0.078   0.020     0.021   0.006     0.129   0.010   < 0.088             0.136   0.030     ---
7135.78     [ArII]  < 0.150             0.115   0.050     0.100   0.019     0.138   0.010     0.118   0.030     0.114   0.030     ---
7319         [OII]                                        0.040   0.007     0.072   0.010                                   
7330         [OII]                                        0.033   0.007     0.087   0.012                     : 0.083   0.040     ---
9068.60     [SIII]  < 0.500           < 0.300             0.207   0.020     0.141   0.030     0.127   0.030   : 0.159   0.070     ---
9529.78     [SIII]  < 0.500           < 0.300             0.472   0.030     0.389   0.080     0.268   0.060   : 0.518   0.200     ---
                                                                                                                                      
c(Hb)                 0.00              0.00              0.00              0.09              0.17              0.00              0.00
log F(Hb)             -16.268           -16.260           -14.937           -15.710          -16.220            -15.733           -16.033
   
\end{verbatim}
\end{document}